\documentclass[aps,twocolumn,prc,superscriptaddress,showpacs,nofootinbib,floatfix,amssymb,amsfonts,amsmath]{revtex4-1}
\usepackage{graphicx}% Include figure files
\usepackage{dcolumn}% Align table columns on decimal point
\usepackage{bm}% bold math
\usepackage{amsmath}    % need for subequations
\usepackage{amsfonts}   %note how statements can be commented out
\usepackage{amssymb}
\usepackage{graphicx}   % for figures

\begin{document}
\title{Covariant density functional theory: Reexamining the structure of
    superheavy nuclei.}
%Covariant density functional theory: the structure of superheavy
%nuclei revisited.}
% with an assessment of theoretical uncertainties.}

\author{S.\ E.\ Agbemava}
\affiliation{Department of Physics and Astronomy, Mississippi
State University, MS 39762}

\author{A.\ V.\ Afanasjev}
\affiliation{Department of Physics and Astronomy, Mississippi
State University, MS 39762}
\affiliation{Center for Computational Sciences, University of Tsukuba, Tsukuba
305-8577, Japan }

\author{T.\ Nakatsukasa}
\affiliation{Center for Computational Sciences, University of Tsukuba, Tsukuba
305-8577, Japan }
\affiliation{RIKEN Nishina Center, Wako 351-0198, Japan}

\author{P. Ring}
\affiliation{Fakult\"at f\"ur Physik, Technische Universit\"at M\"unchen,
 D-85748 Garching, Germany}

\date{\today}

\begin{abstract}
 A systematic investigation of even-even superheavy elements in the region of
proton numbers $100 \leq Z \leq 130$ and in the region of neutron numbers
from the proton-drip line up to neutron number $N=196$
is presented. For this study we use five most up-to-date covariant energy
density functionals of different types, with a non-linear meson coupling, with
density dependent meson couplings, and with density-dependent zero-range
interactions. Pairing correlations are treated within relativistic
Hartree-Bogoliubov (RHB) theory based on an effective separable
particle-particle interaction of finite range and deformation effects are
taken into account. This allows us to assess the spread of theoretical
predictions within the present covariant models
for the binding energies,  deformation parameters, shell structures and
$\alpha$-decay half-lives.  Contrary to the previous studies in covariant
density functional theory, it was found that the impact of $N=172$ spherical
shell gap on the structure of superheavy elements is very limited. Similar
to non-relativistic functionals some covariant functionals predict the
important role played by the spherical $N=184$ gap.  For these functionals
(NL3*, DD-ME2 and PC-PK1), there is a band of spherical nuclei along and
near the $Z=120$ and $N=184$ lines. However, for other functionals (DD-PC1
and DD-ME$\delta$) oblate shapes dominate at and in the vicinity of these
lines. Available experimental data are in general described with comparable
accuracy and do not allow to discriminate these predictions.
\end{abstract}


\maketitle

%%%%%%%%%%%%%%%%%%%%%%%%%%%%%%%%%%%
\section{Introduction}
%%%%%%%%%%%%%%%%%%%%%%%%%%%%%%%%%%%

Science is driven by the efforts to understand unknowns. In
low-energy nuclear physics many of such unknowns are located at the
extremes of the nuclear landscape \cite{Eet.12,AARR.13,AARR.14}. The region
of superheavy elements (SHE), characterized by the extreme values of
proton number $Z$, is one of such extremes. Contrary to other regions
of the nuclear chart, the SHEs are stabilized only by quantum shell effects.
Because of this attractive feature and the desire to extend the nuclear
landscape to higher $Z$ values, this region is an arena of active
experimental and theoretical studies.

Currently available experimental data reach proton number
$Z=118$ \cite{Z=116-118-year2006,Z=117-118-year2012} and dedicated
experimental facilities such as the Dubna Superheavy Element Factory will
hopefully allow to extend the region of SHEs up to $Z=120$ and
for a wider range on neutron numbers at lower $Z$ values. Unfortunately,
even this facility will not be able to reach the predicted centers of the
island of stability of SHEs at $(Z=114, N=184)$, $(Z=120, N=172/184)$ and
$(Z=126, N=184)$ as given by microscopic+macroscopic (MM) approaches
\cite{NNSSWGM.68,NTSZWGLMN.69,PS.91,CDHMN.96,MN.94} or by covariant
\cite{Rutz_PhysRevC.56.238.1997,BRRMG.99,A250,ZMZGT.05} and Skyrme
\cite{CDHMN.96,BRRMG.99} density functional theories (DFT),
respectively.

 One has to recognize that the majority of systematic DFT studies
of the shell structure of SHEs has been performed in the 90ies of the
last century and at the beginning of the last decade. These studies
indicate that the physics of SHEs is much richer in the DFT framework
than in MM approaches. This is due to self-consistency effects which
are absent in the MM approaches. For example, they manifest themselves
by a central depression in the density distribution of spherical SHEs
\cite{BRRMG.99,AF.05-dep} which has not been seen in the MM approaches.
Moreover, besides the successful covariant energy
density functionals (CEDFs) NL1 \cite{NL1} and NL3 \cite{NL3},
during the last ten years a new generation of energy density
functionals has been developed both in covariant
\cite{DD-ME2,DD-PC1,NL3*,PC-PK1,DD-MEdelta} and in non-relativistic
\cite{UNEDF2,GHGP.09} frameworks; they are characterized by an improved
global performance \cite{UNEDF2,GHGP.09,AARR.14}. In addition, the experimental
data on SHEs became much richer \cite{HG.08,OU.15} in these years.

  In such a situation it is necessary to reanalyze the structure
of superheavy nuclei using both the full set of available experimental
data on SHEs and the new generation of energy density functionals. There
are several goals of this study. First, we will investigate the accuracy of the
description of known SHEs with the new generation of covariant energy
density functionals and find whether the analysis of existing experimental
data allows to distinguish the predictions of different functionals for nuclei
beyond the known region of SHEs. Second, the comparative analysis of the results
obtained with several state-of-the-art functionals will allow to
estimate the spreads of theoretical predictions when extending the region of SHEs
beyond the presently known, to establish their major sources and to define the
physical observables and regions of SHEs which are less affected by these spreads.

This is a very ambitious goal and, as a consequence, several restrictions
are imposed. This study is performed only in the framework of covariant
density functional  theory (CDFT) \cite{VALR.05}. We use the covariant energy
density functionals (CEDF) NL3* \cite{NL3*}, DD-ME2 \cite{DD-ME2},
DD-ME$\delta$ \cite{DD-MEdelta}, DD-PC1
\cite{DD-PC1} and PC-PK1 \cite{PC-PK1}. They are the state-of-the-art
functionals representing the major
classes of CDFTs (for more details see the discussion in Sect. II of
Ref.\ \cite{AARR.14})\footnote{Note that the functional PC-PK1 has
not been used and discussed in Ref.\ \cite{AARR.14} because
global studies with it have been performed by
the Peking group in Ref.\ \cite{ZNLYM.14}. It is a state-of-the-art
functional for point coupling models with cubic and quartic interactions of zero
range \cite{PC-PK1}.}.  Moreover, their global performance
has recently been analyzed in Refs.\ \cite{AARR.14,ZNLYM.14} and they are
characterized by an improved description of experimental data as compared with
previous generation of CEDFs. Moreover, the study of Refs.\ \cite{AARR.14,AARR.15}
provides theoretical spreads in the description of known nuclei and
their propagation towards the neutron-drip line obtained with four CEDFs (NL3*,
DD-ME2, DD-ME$\delta$ and DD-PC1). This is in contrast with earlier studies
of SHEs in the CDFT framework based on the functionals whose performance
was tested only in very restricted regions of the nuclear chart and for which
no analysis of theoretical spreads has been carried out.

Contrary to  many
earlier CDFT studies of the shell structure in superheavy nuclei
restricted to spherical shape, in this investigation the effects
of deformation are taken into account. As it will be shown later,
neglecting deformation can lead to wrong conclusions since many
SHEs are
characterized by soft potential energy surfaces with coexisting minima.
Considering the numerical complexity of global investigations we restrict
our calculations to axial reflection
symmetric shapes. Such calculations are realistic for the absolute majority of
the ground states. Octupole deformation does not play a role in the
ground states of SHEs \cite{PNLV.12,AAR.14} but it affects the properties
of the outer fission barriers \cite{AAR.12,PNLV.12}. In the current manuscript
those are not considered in detail  (see the discussion in Sect.\ \ref{Out_fission}).
Although triaxiality may play a role in the ground states of some SHEs \cite{CHN.05},
such cases are rather exceptions than the rule \cite{AAR.12,CHN.05,PNV.13,PNLV.12}.
Moreover, model predictions for stable triaxial deformation in the
ground states vary drastically between the models, even in
experimentally known nuclei \cite{MBCOISI.08}, and are frequently not supported
by the analysis of experimental data \cite{Mo-Ru.13}. In addition, triaxial RHB
\cite{CRHB} calculations are at present too time-consuming to be undertaken
on a global scale. These arguments justify neglecting of triaxiality in the
description of the ground states in this investigation.

In addition we restrict our investigation to even-even nuclei. Unfortunately,
no reliable configuration assignments exist for ground states of experimentally
known odd-mass SHEs to be confronted with the theory. However, systematic
studies of the accuracy of the reproduction of energies of deformed
one-quasiparticle states in actinides are available for the CEDF  NL3*
\cite{AS.11,DABRS.15}.

  The paper is organized as follows. Section \ref{theory_details}
describes the details of the calculations. The single-particle
structure and shell gaps at spherical shape together with
the spreads in their description are discussed in Sec.\
\ref{sp-states}.  The impact of deformation on the properties
of SHEs is considered in Sect.\ \ref{def-impact}. Section \ref{def-system}
contains the systematics of calculated charge quadrupole deformations.
The validity of the $\delta_{2n}$ and $\delta_{2p}$ quantities
as the indicators of shell gaps is discussed in Sec.\ \ref{Sect-delta_2}.
We report on masses and separation energies in Sec.\ \ref{Sep-energies}.
The $\alpha$-decay properties are considered in Sec.\ \ref{Alpha-decay}.
Finally, Sec.\ \ref{concl} summarizes the results of our work.

%%%%%%%%%%%%%%%%%%%%%%%%%%%%%%%%%%%%%%%%%%%%%%%%%
\section{The details of the theoretical calculations}
\label{theory_details}
%%%%%%%%%%%%%%%%%%%%%%%%%%%%%%%%%%%%%%%%%%%%%%%%%

%%%%%%%%%%%%%%%%%%%%%%%%%%%%%%%%%%%%%%%%%%%%%%%%%%%%%%%%%%%%%%
\begin{figure*}[ht]
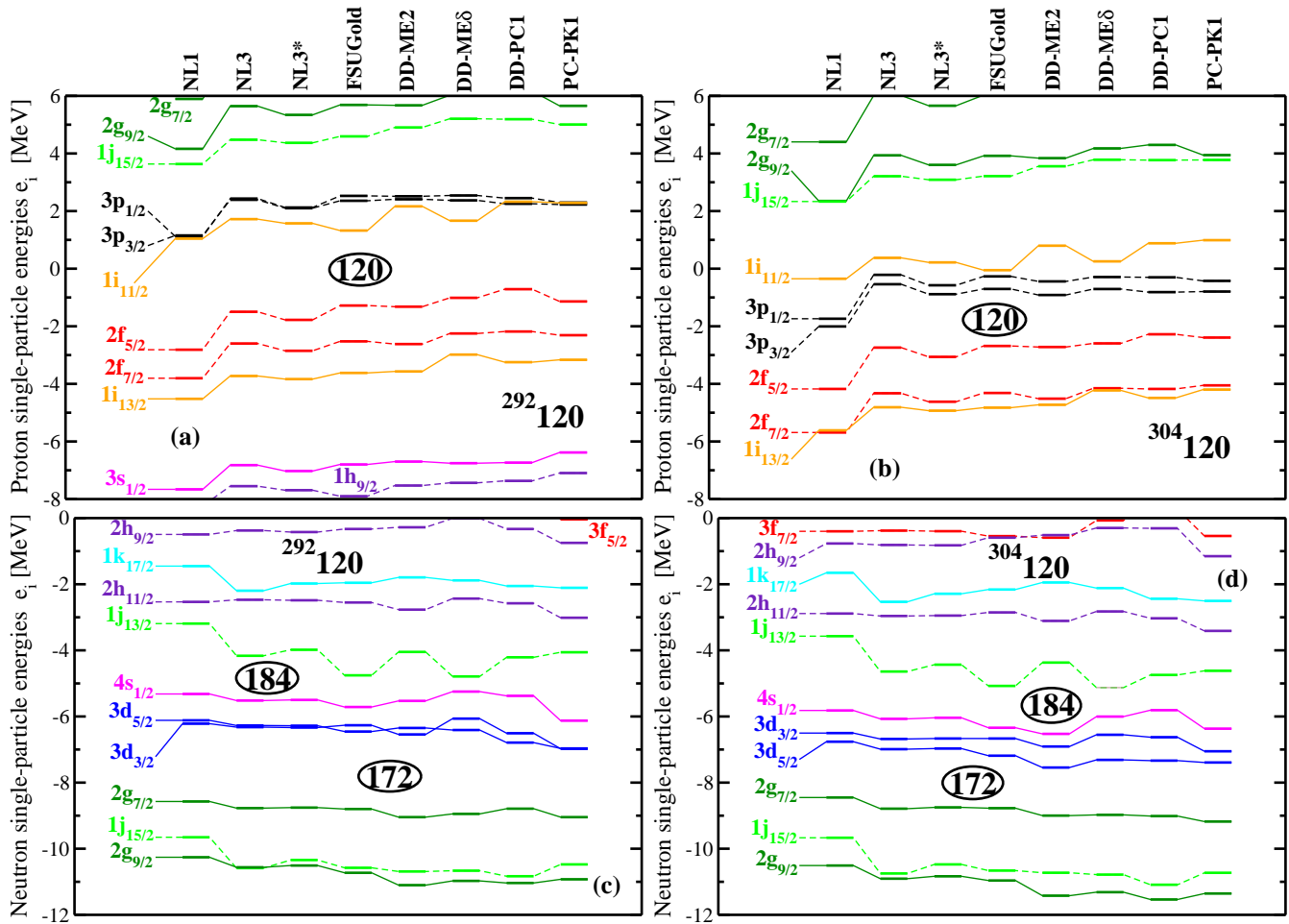

\includegraphics[angle=0,width=8.8cm]{fig-1-a.eps}
\includegraphics[angle=0,width=8.8cm]{fig-1-b.eps}
\includegraphics[angle=0,width=8.8cm]{fig-1-c.eps}
\includegraphics[angle=0,width=8.8cm]{fig-1-d.eps}
\caption{(Color online) Neutron (left panels) and proton (right panels)
single-particle states at spherical shape in the SHEs
$^{292}$120 and $^{304}$120. They are determined with the
indicated CEDFs in RMF calculations without pairing.
Solid and dashed connecting lines are used for positive and
negative parity states. Spherical gaps are indicated.
The different CEDFs have been arranged in such a way that we first start
from the functionals with a non-linear $\sigma$ coupling
(NL1 \cite{NL1}, NL3 \cite{NL3}, NL3* \cite{NL3*}), i. e. with a density dependence in the isoscalar channel,
then we continue with the functional FSUGold \cite{FSUGold} which has in addition
a non-linear coupling between $\omega$ and $\rho$ mesons
and therefore also a density dependence in the isovector channel, then we plot
the results for the functionals with explicit density dependent
meson-nucleon couplings in all channels (DD-ME2 \cite{DD-ME2} and DD-ME$\delta$ \cite{DD-MEdelta})
and finally we end with the point coupling functionals
(DD-PC1 \cite{DD-PC1} and PC-PK1 \cite{PC-PK1}).}
\label{spectra}
\end{figure*}
%%%%%%%%%%%%%%%%%%%%%%%%%%%%%%%%%%%%%%%%%%%%%%%%%%%%%%%%%%%%%%%%%

%%%%%%%%%%%%%%%%%%%%%%%%%%%%%%%%%%%%%%%%%%%%%%%%%%%%%%%%%%%%%%
\begin{figure}[ht]
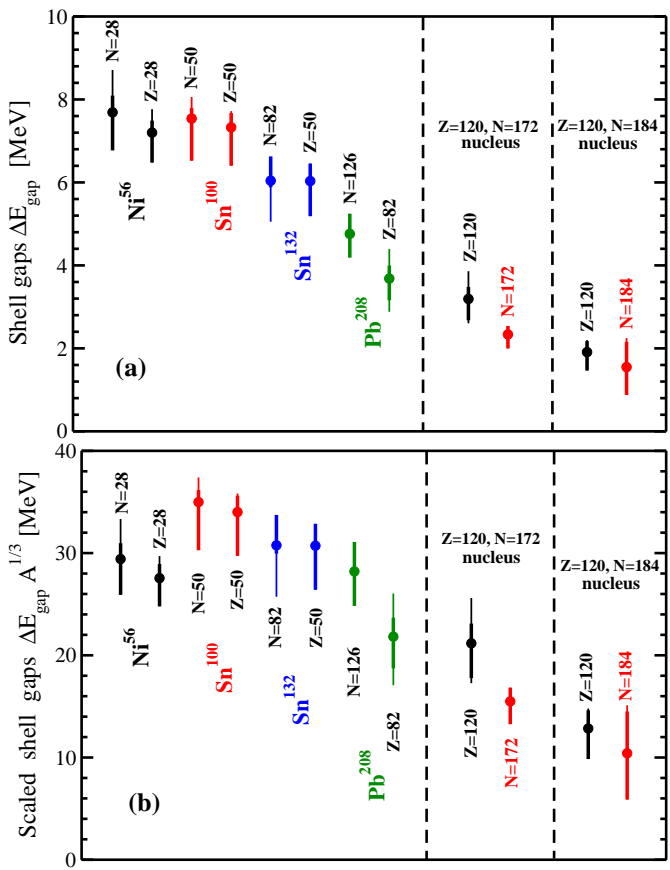

\centering
\includegraphics[angle=0,width=8.8cm]{fig-2-a.eps}
\includegraphics[angle=0,width=8.8cm]{fig-2-b.eps}
\caption{(Color online) (a) Neutron and proton shell
gaps $\Delta E_{\rm gap}$ of the indicated nuclei. The average
(among ten used CEDFs) size of the shell gap is shown by a
solid circle. Thin and thick vertical lines are used to show
the spread of the sizes of the calculated shell gaps; the top
and bottom of these lines
corresponds to the largest and smallest shell gaps amongst
the considered set of CEDFs. Thin lines show this spread for
all employed CEDFs, while thick lines are used for the subset
of four CEDFs (NL3*, DD-ME2,  DD-ME$\delta$ and DD-PC1).
Particle numbers corresponding to the shell gaps are indicated.
(b) The same as in panel (a) but with the sizes of the
shell gaps and the spreads in their predictions scaled with
mass factor $A^{1/3}$.}
\label{gap-sizes}
\end{figure}
%%%%%%%%%%%%%%%%%%%%%%%%%%%%%%%%%%%%%%%%%%%%%%%%%%%%%%%%%%%%%

In the present manuscript, the RHB framework is used for
systematic studies of all $Z=96-130$ even-even actinides and
SHEs from the proton-drip line up to neutron number $N=196$. The details of
this formalism have been discussed in Secs.\ II-IV of Ref.\
\cite{AARR.14} and Sec.\ II of Ref.\ \cite{AARR.15}. Thus,
we only provide a general outline of the features specific for
the current RHB calculations.

We consider only axial and parity-conserving intrinsic
states and solve the RHB-equations in an axially deformed
oscillator basis \cite{Koepf1988_PLB212-397,GRT.90,RGL.97,Niksic2014_CPC185-1808}.
The truncation of the basis is  performed in such a way
that all states belonging to the shells up to $N_F = 20$
fermionic shells and $N_B = 20$ bosonic shells are taken into
account.  As tested in a number of calculations with $N_F=26$
and $N_B=26$ for heavy neutron-rich nuclei, this truncation
scheme provides sufficient numerical accuracy.  For each nucleus
the potential energy curve is obtained in a large deformation
range from $\beta_2=-1.0$ up to $\beta_2=1.05$ in steps of
$\beta_2=0.02$ by means of a constraint on the quadrupole
moment $Q_{20}$. Then, the correct ground state configuration
and its energy are defined; this procedure  is especially
important for the cases of shape coexistence.

In order to avoid the uncertainties connected with the definition of
the size of the pairing window, we use the separable form of the
finite range Gogny pairing interaction introduced by Tian et al \cite{TMR.09}. Its matrix
elements in $r$-space have the form
\begin{eqnarray}
\label{Eq:TMR}
V({\bm r}_1,{\bm r}_2,{\bm r}_1',{\bm r}_2') &=& \nonumber \\
= - G \delta({\bm R}-&\bm{R'}&)P(r) P(r') \frac{1}{2}(1-P^{\sigma})
\label{TMR}
\end{eqnarray}
with ${\bm R}=({\bm r}_1+{\bm r}_2)/2$ and ${\bm r}={\bm r}_1-{\bm r}_2$
being the center of mass and relative coordinates. The form factor
$P(r)$ is of Gaussian shape
\begin{eqnarray}
P(r)=\frac{1}{(4 \pi a^2)^{3/2}}e^{-r^2/4a^2}
\end{eqnarray}
The two parameters $G=738$ fm$^3$ and $a=0.636$ fm of this interaction
are the same for protons and neutrons and have been derived in Ref.\
\cite{TMR.09} by a mapping of the $^1$S$_0$ pairing gap of infinite
nuclear matter to that of the Gogny force D1S~\cite{D1S}.

  As follows from the RHB studies with the CEDF NL3* of odd-even mass
staggerings, moments of inertia and pairing gaps the Gogny D1S
pairing and its separable form (Eq. (\ref{Eq:TMR})) work well in
actinides (Refs.\ \cite{AO.13,AARR.14,DABRS.15}). The weak dependence
of its pairing  strength on the CEDF has been seen in the studies of
pairing and rotational properties of actinides in Refs.\ \cite{A250,AO.13},
of pairing gaps in spherical nuclei in Ref.\ \cite{AARR.14} and
of pairing energies in Ref.\ \cite{AARR.15}. Thus, the same pairing
(Eq. (\ref{Eq:TMR})) is used also in the calculations with DD-PC1, DD-ME2,
DD-ME$\delta$, and PC-PK1. Considering the global character of this
study, this is a reasonable choice.

Any extrapolation beyond the known region requires some estimate
of theoretical uncertainties. This issue has been discussed in the
details in Refs.\ \cite{RN.10,DNR.14} and in the context of global
studies  within CDFT in the introduction of Ref.\ \cite{AARR.14}. In
the present manuscript, we concentrate on the uncertainties related
to the present choice of energy density functionals which can be
relatively easily deduced globally \cite{AARR.14}. We therefore
define spreads of theoretical predictions for a given physical 
observable as \cite{AARR.14}
\begin{equation}
O(Z,N) = |O_{max}(Z,N) - O_{min}(Z,N)|,
\end{equation}
where $O_{max}(Z,N)$ and $O_{min}(Z,N)$ are the largest and smallest
values of the physical observable $O(Z,N)$ obtained with the
employed set of CEDFs for the $(Z,N)$ nucleus. Note that these
spreads are only a crude approximation to the {\it systematic}
theoretical errors discussed in Ref.\ \cite{DNR.14} since they
are due to a very small number of functionals which do not form
an independent statistical ensemble. Despite this fact they provide
an understanding which observables/aspects of many-body physics
can be predicted with a higher level of confidence than others for
density functionals of the given type.
Moreover, it is expected that they will indicate which aspects
of many-body problem have to be addressed with more care during
the development of next generation of EDFs.

%%%%%%%%%%%%%%%%%%%%%%%%%%%%%%%%%%%%%%%%%%%%%%%%%%%%%%%%
\section{Single-particle structures at spherical shape}
\label{sp-states}
%%%%%%%%%%%%%%%%%%%%%%%%%%%%%%%%%%%%%%%%%%%%%%%%%%%%%%%%

 As discussed in the introduction, superheavy nuclei are stabilized by shell
effects, i.e. by a large shell gap or at least a considerably reduced density
of the single-particle states.
Therefore, since a long time an island of stability has been
predicted in the CDFT for very heavy nuclei in the region around
the proton number $Z=120$
\cite{Rutz_PhysRevC.56.238.1997,BRRMG.99,A250,ZMZGT.05}.
Fig.\ \ref{spectra} shows neutron and proton
single-particle spectra of the nuclei $^{292}$120 and $^{304}$120 obtained in
spherical relativistic mean field (RMF) calculations. Similar figures have
been given for the functionals NL3 and DD-ME2 in Refs.\ \cite{BRRMG.99,LG.14}.
Note that a detailed comparison of several other covariant and non-relativistic Skyrme
functionals is presented in Ref.\ \cite{BRRMG.99}. In Fig.\ \ref{spectra}
we show the results for an extended set of 8 CEDFs.
The global performance of the
CEDFs NL3*, DD-ME2, DD-ME$\delta$ and DD-PC1 has been studied in Ref.\ \cite{AARR.14}. The shell
gaps at  $Z=120$ and at $N=172$  are especially pronounced in the nucleus
$^{292}$120 (left panels of Fig.\ \ref{spectra}). This is a consequence of the
presence of a central depression in the density distribution generated by a
predominant occupation of the high-$j$ orbitals above the occupied single-particle
states in $^{208}$Pb. Because of their large $\ell$-values these orbitals produce
density at the surface of the nucleus.
Filling up the low-$j$ neutron orbitals above the Fermi surface
of the $^{292}$120 nucleus on going from $N=172$ up to $N=184$  leads
to a flatter density distribution in the $N=184$ system
\cite{AF.05-dep} . As a consequence, the $Z=120$ and $N=172$ shell gaps are
reduced and $N=184$ gap is increased (right panels of Fig.\ \ref{spectra}).
As one can see in Fig.\ \ref{spectra}, these features are rather general
and do not depend much on the specific density functional.

  Of course, as shown in Fig.\ \ref{spectra}, there are theoretical
uncertainties in the description of the single-particle energies and in
their relative positions. The precise size of the large shell gaps depend
on the functional. The corresponding spreads are summarized in Fig.\ 
\ref{gap-sizes}a, which shows the average sizes of these shell gaps and 
the spreads in their predictions. These gaps in the superheavy region are 
compared with the calculated gaps in lighter doubly magic nuclei, such as 
$^{56}$Ni, $^{100}$Sn, $^{132}$Sn and $^{208}$Pb.

 Since the nuclear radius $R\approx r_{0}A^{1/3}$, i.e. the average width of
the potential, increases with the mass number $A$, the shell gaps decrease
with $A^{-1/3}$ and by this reason we show in Fig.\ \ref{gap-sizes}b
the shell gaps and their spreads scaled with a factor $A^{1/3}$.
These scaled shell gaps are considerably more constant with $A$, but there is still
a tendency that even the scaled gaps decrease in general with $A$. This is
probably related to the spin-orbit coupling, which is proportional to the
orbital angular  momentum $\ell$, since it causes an increasing downward shift of the
high-$j$ intruder levels. The spreads give some information on the
theoretical uncertainties of the sizes of the calculated gaps. Definitely, the impact of these
spreads on the model predictions depends on the ratio of their size
with respect of the size of calculated shell gaps.
The presence of theoretical spreads has less severe consequences on the
predictions of spherical nuclei around magic gaps at $Z=28,50,82$ and $N=28,50,82,126$
than on similar predictions for SHEs since the former typically have larger
shell gaps for comparable theoretical uncertainties. Of course, this is only
true in general. The $N=172$ shell gap in the nucleus $^{292}$120 forms
an exception. It is more or less the same for all the CEDFs under consideration
and therefore its uncertainty deduced from these spreads is relatively small.

It is evident that the predictive power for new shell gaps in the superheavy
region depends on the quality of the description of the single-particle
energies of the various CEDFs. Considering Figs.\ \ref{spectra} and
\ref{gap-sizes} one can hope that an improvement in the DFT description of
the energies of the single-particle states in known nuclei will also reduce
the uncertainties in the prediction of the shell structure of the SHEs. It is
well known that in DFT the single-particle energies are auxiliary quantities
and there are problems in their precise description within this
framework. It is generally assumed that this has two reasons. First
the coupling of the single-particle motion to low-lying surface vibrations
has to be taken into account, and second there is not enough known about
the influence of additional terms in the Lagrangian, such as tensor forces~\cite{LKSOR.09}.
Particle-vibrational coupling is particularly large in spherical nuclei.
So far, its influence on the accuracy of the description of the single-particle energies
and on the sizes of shell gaps has been studied in relativistic particle-vibration
\cite{Litvinova2006_PRC73-044328,LA.11} and quasiparticle-vibration
\cite{AL.15} coupling models with the CEDFs NL3
\cite{NL3} and NL3* \cite{NL3*} only. The experimentally known gaps of $^{56}$Ni,
$^{132}$Sn and $^{208}$Pb are reasonably well described in the
relativistic particle-vibration calculations of Ref.\ \cite{LA.11}. Also, the
impact of particle-vibration coupling on spherical shell gaps in SHEs has
been investigated in Refs.\ \cite{LA.11,L.12}. Although this effect, in
general, decreases the size of shell gaps, the $Z=120$ gap still remains
reasonably large but there is a competition between the smaller $N=172$ and
$N=184$ gaps. The accuracy of the description of the energies of
one-quasiparticle deformed states in the rare-earth and actinide region has
been statistically evaluated in Ref.\ \cite{AS.11} within the framework of
relativistic Hartree-Bogoliubov theory. On the one hand, these studies have
proven some success of CDFT: the covariant functionals provide a reasonable
description of the single-particle properties despite the fact that such
observables were not used in their fit. On the other hand, they illustrate the
need for a better description of the single-particle energies.

%%%%%%%%%%%%%%%%%%%%%%%%%%%%%%%%%%%%%%%%%%%%%%%%%%%%%%%%%%%%%%%%%%%%%%%%%%%%%
\section{The impact of deformation on the properties of superheavy nuclei.}
\label{def-impact}
%%%%%%%%%%%%%%%%%%%%%%%%%%%%%%%%%%%%%%%%%%%%%%%%%%%%%%%%%%%%%%%%%%%%%%%%%%%%%

Although it is commonly accepted that the large spherical
shell gaps at $Z=120$ and $N=172$ define the center of the island of stability
of SHEs for the majority of the covariant functionals \cite{BRRMG.99,A250},
these conclusions were mostly obtained in investigations restricted to
spherical shapes. In addition, some calculations suggest \cite{LG.14,BNR.01},
or do not exclude \cite{A250}, the existence of a spherical shell gap at the
neutron number $N=184$. However, as discussed below, the inclusion of
deformation can change the situation drastically for some functionals.

\begin{figure*}[ht]
\centering
\includegraphics[angle=0,width=17cm]{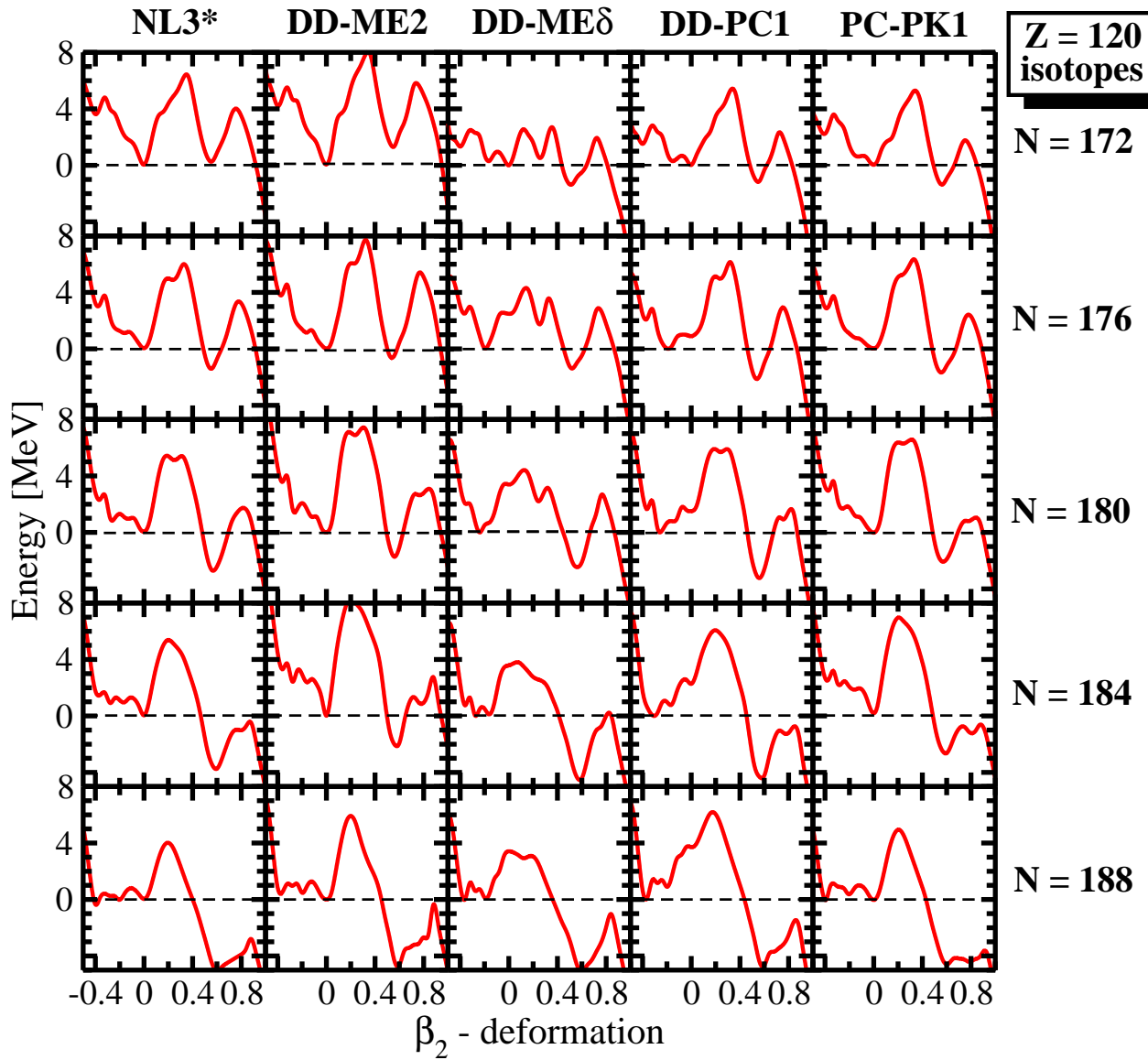}
\caption{ (Color online) Deformation energy curves for the chain
of $Z=120$ isotopes obtained in axial RHB calculations
with the indicated CEDFs. The energy of the spherical or oblate ground states are set to
zero.}
\label{Z=120-pes}
\end{figure*}

\begin{figure*}[ht]
\centering
\includegraphics[angle=0,width=17.5cm]{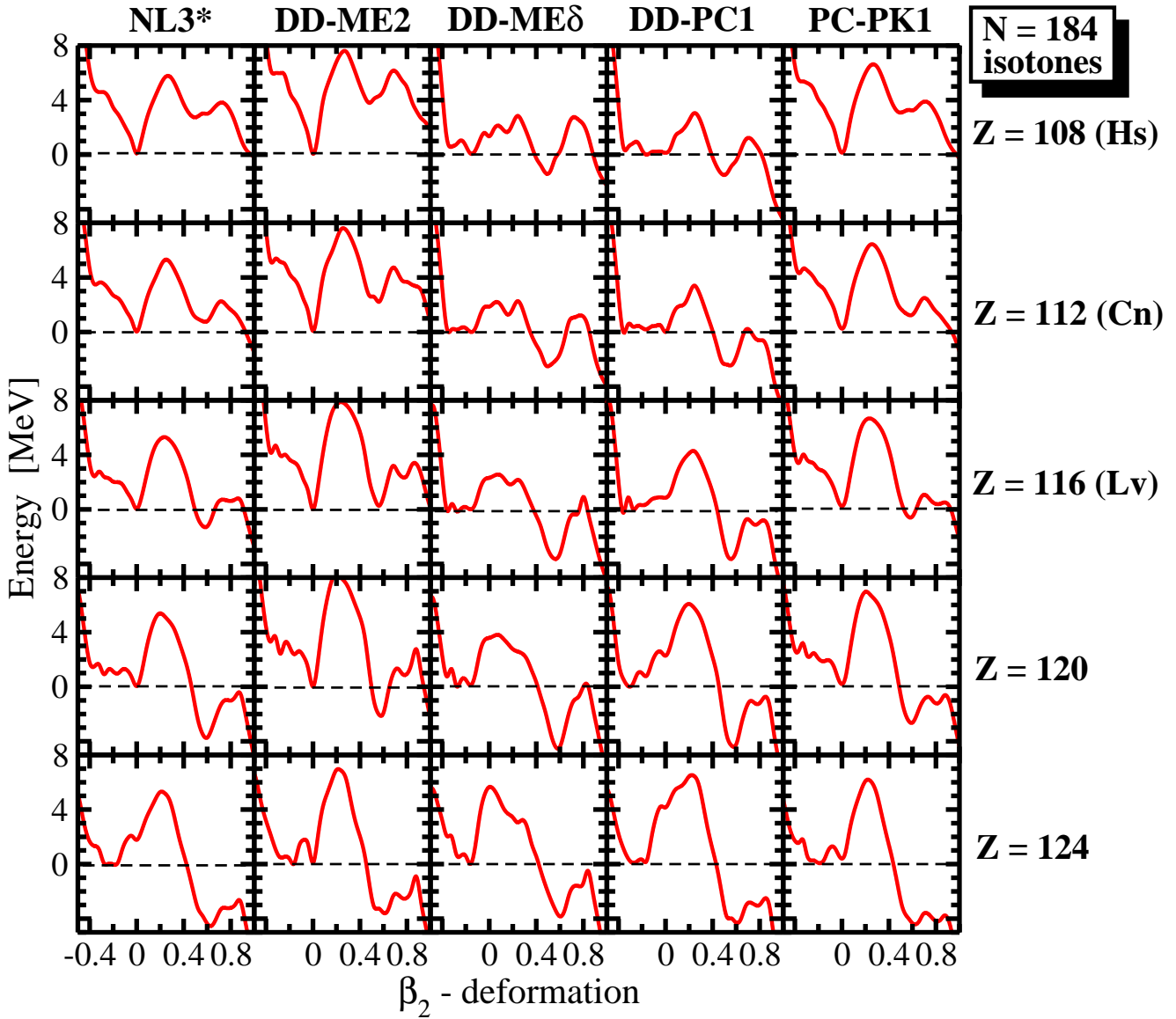}
\caption{(Color online) The same as in Fig.\ \protect\ref{N=184-pes}
but for $N=184$ isotones.}
\label{N=184-pes}
\end{figure*}

To illustrate this fact, the deformation energy curves of the $Z=120$ isotopes and
the $N=184$ isotones are presented in Figs.\ \ref{Z=120-pes} and \ref{N=184-pes}.
Here we restrict our considerations to five CEDFs, namely,
NL3*, DD-ME2, DD-ME$\delta$, DD-PC1 and PC-PK1, whose global
performance is well established \cite{AARR.14,ZNLYM.14}.  In the following
discussion we neglect the prolate superdeformed minimum, which is sometimes even
lower than the spherical or oblate minimum, because of the reasons discussed in
detail in Sec. \ref{Out_fission}.
In Figs.\ \ref{Z=120-pes} and \ref{N=184-pes} the lowest spherical
or oblate minimum is considered as the ground state and indicated by a dashed
horizonal line. In Fig.\ \ref{Z=120-pes}
we see that the ground states of the $Z=120$ isotopes with $N=172-184$
are spherical for NL3*, DD-ME2, and PC-PK1.
This is a consequence of the presence of the large $Z=120$ spherical shell
gap (see Fig.\ \ref{spectra}). For these three functionals, the increase of
neutron number $N$ leads to softer potential energy curves for
$\beta_2$ values between $-0.4$ and $0.0$. As a result, for $N=188$ an oblate
minimum either becomes lowest in energy (for NL3*) or competes in energy with
the spherical solution (for DD-ME2 and PC-PK1). This softness of the potential
energy curves is even more pronounced for the DD-ME$\delta$ and DD-PC1, for
which the oblate solution is lower in energy than the spherical solution in all
displayed nuclei apart from $N=172$ (Fig.\ \ref{Z=120-pes}).

Although it is tempting
to relate this feature to the fact that the size of the $Z=120$ gap is smallest
among the employed functionals for DD-ME$\delta$ in the $(Z=120,N=172)$ nucleus
and for DD-PC1 in the $(Z=120,N=184)$ nucleus (see Fig.\ \ref{spectra}a and b),
this explanation is too simplistic. This is because even for the cases when the
sizes of the $Z=120$ gap are very similar (compare, for example, their sizes
for DD-ME2 and DD-ME$\delta$ in the  $(Z=120,N=184)$ nucleus [Fig.\
\ref{spectra}b]), the deformations of their minima in the ground state are
different. This strongly suggest that the evolution of the single-particle
structure with deformation, which leads to negative shell correction energies at
oblate shape, is responsible for the observed features. Thus, not only the
size of the spherical shell gaps but also the location of the
single-particle states below and above these gaps is responsible for the
observed features.

%%%%%%%%%%%%%%%%%%%%%%%%%%%%%%%%%%%%%%%%%%%%%%%%%%%%%%%%%%%%%
%%%%%%%%%%%%%%%%%%%%%%%%%%%%%%%%%%%%%%%%%%%%%%%%%%%%%%%%%%%%%

%%%%%%%%%%%%%%%%%%%%%%%%%%%%%%%%%%%%%%%%%%%%%%%%%%%%%%%%%%%%%%%%
%%%%%%%%%%%%%%%%%%%%%%%%%%%%%%%%%%%%%%%%%%%%%%%%%%%%%%%%%%%%%

\begin{figure*}[ht]
\includegraphics[angle=0,width=8.8cm]{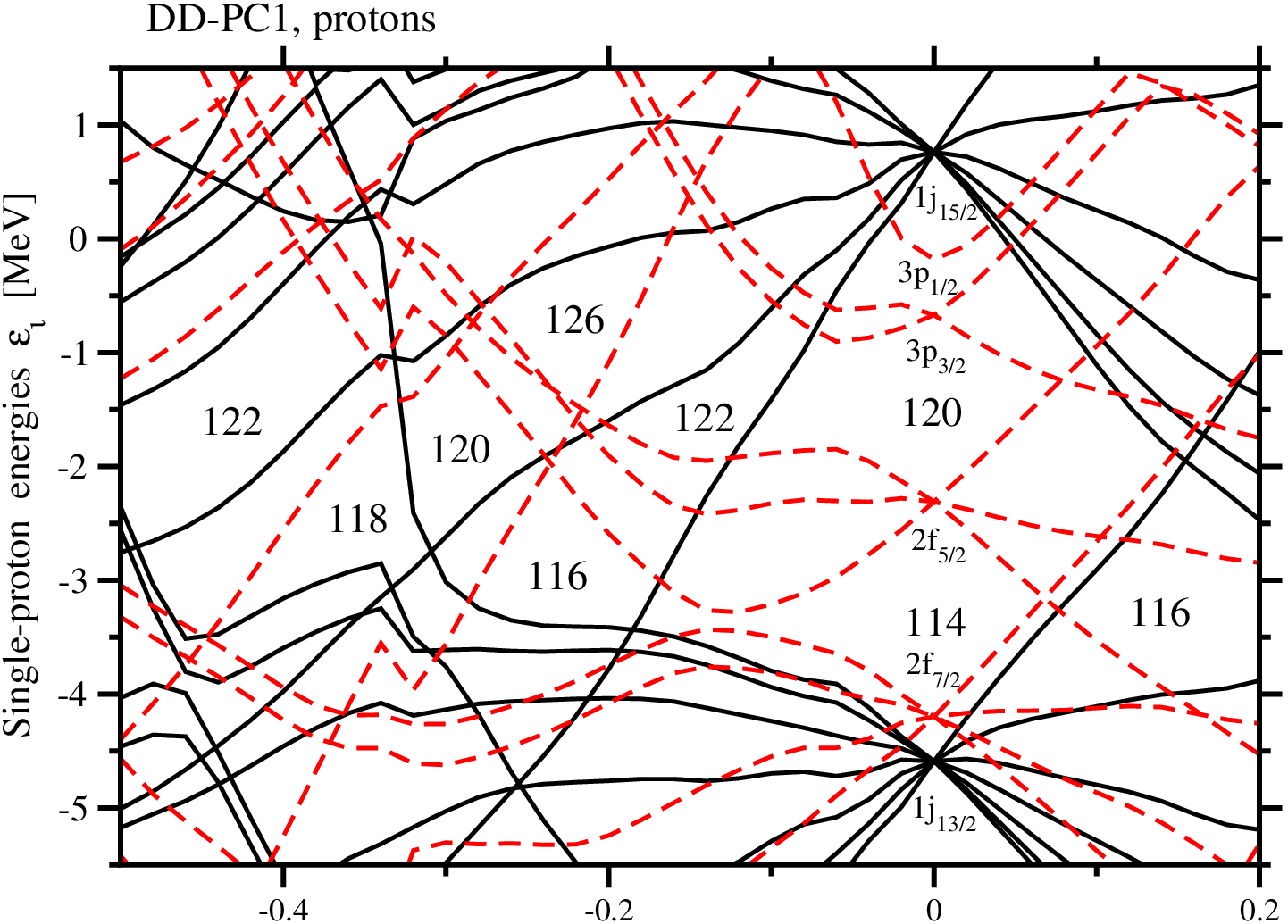}
\includegraphics[angle=0,width=8.8cm]{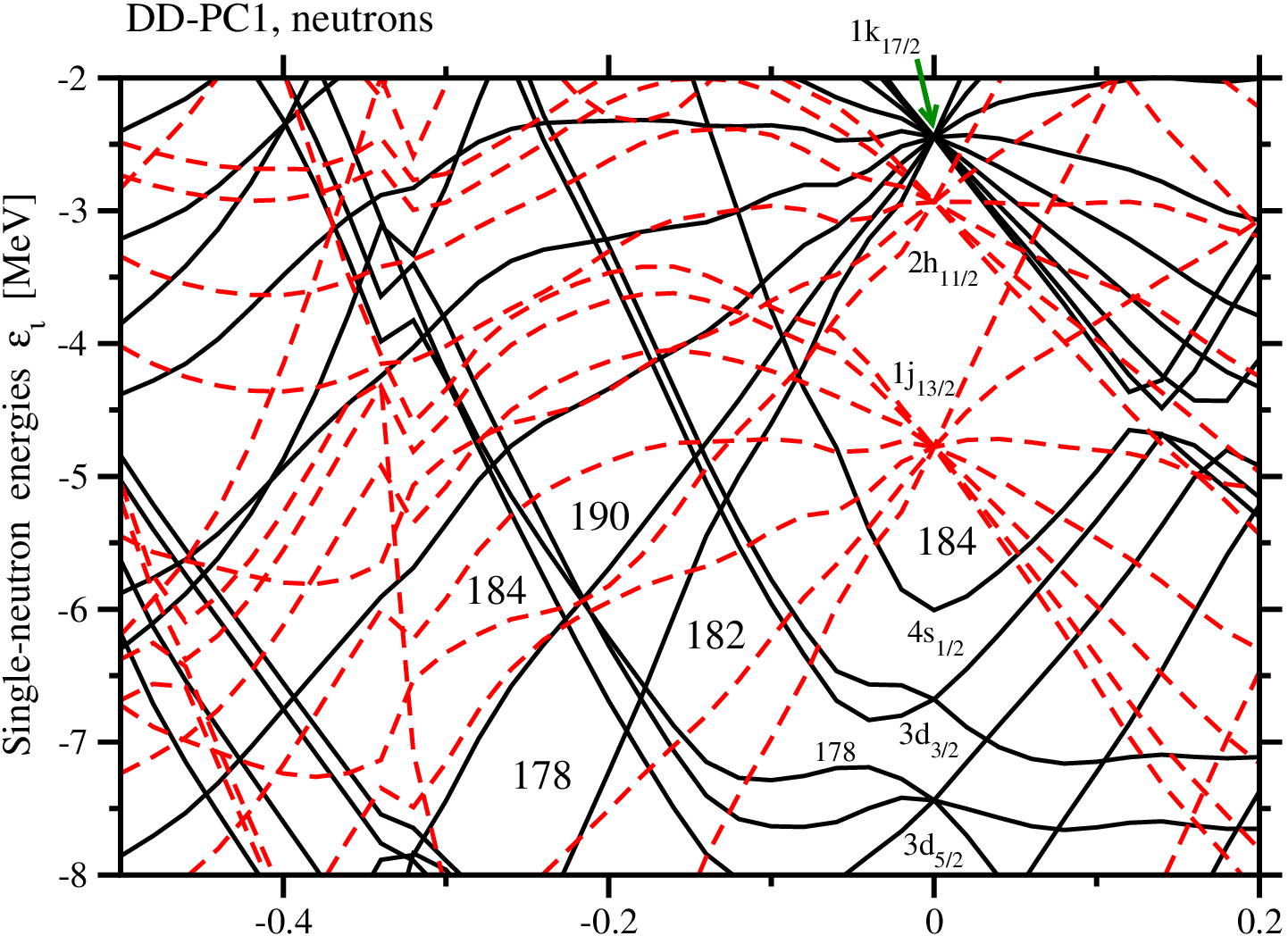}
\includegraphics[angle=0,width=8.8cm]{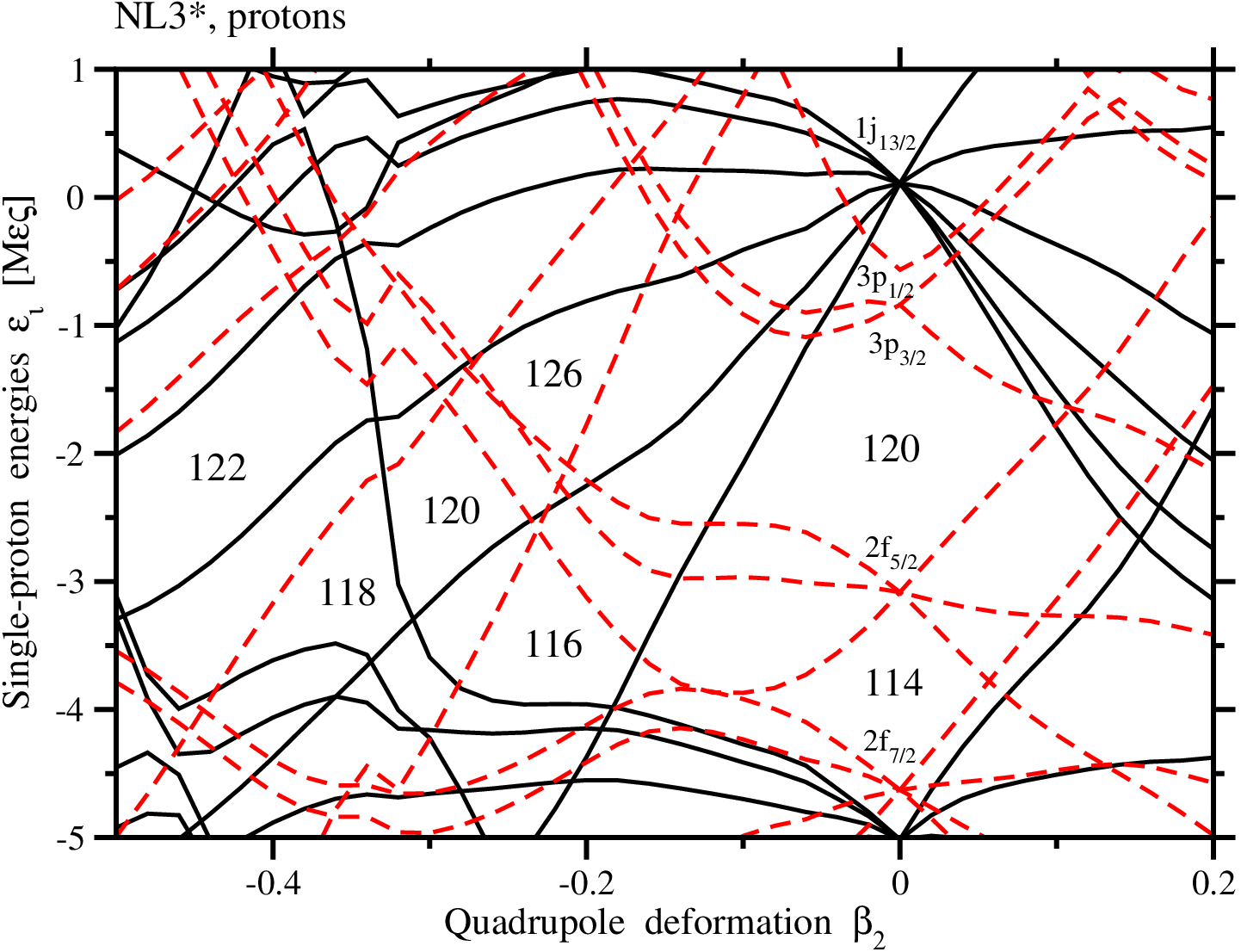}
\includegraphics[angle=0,width=8.8cm]{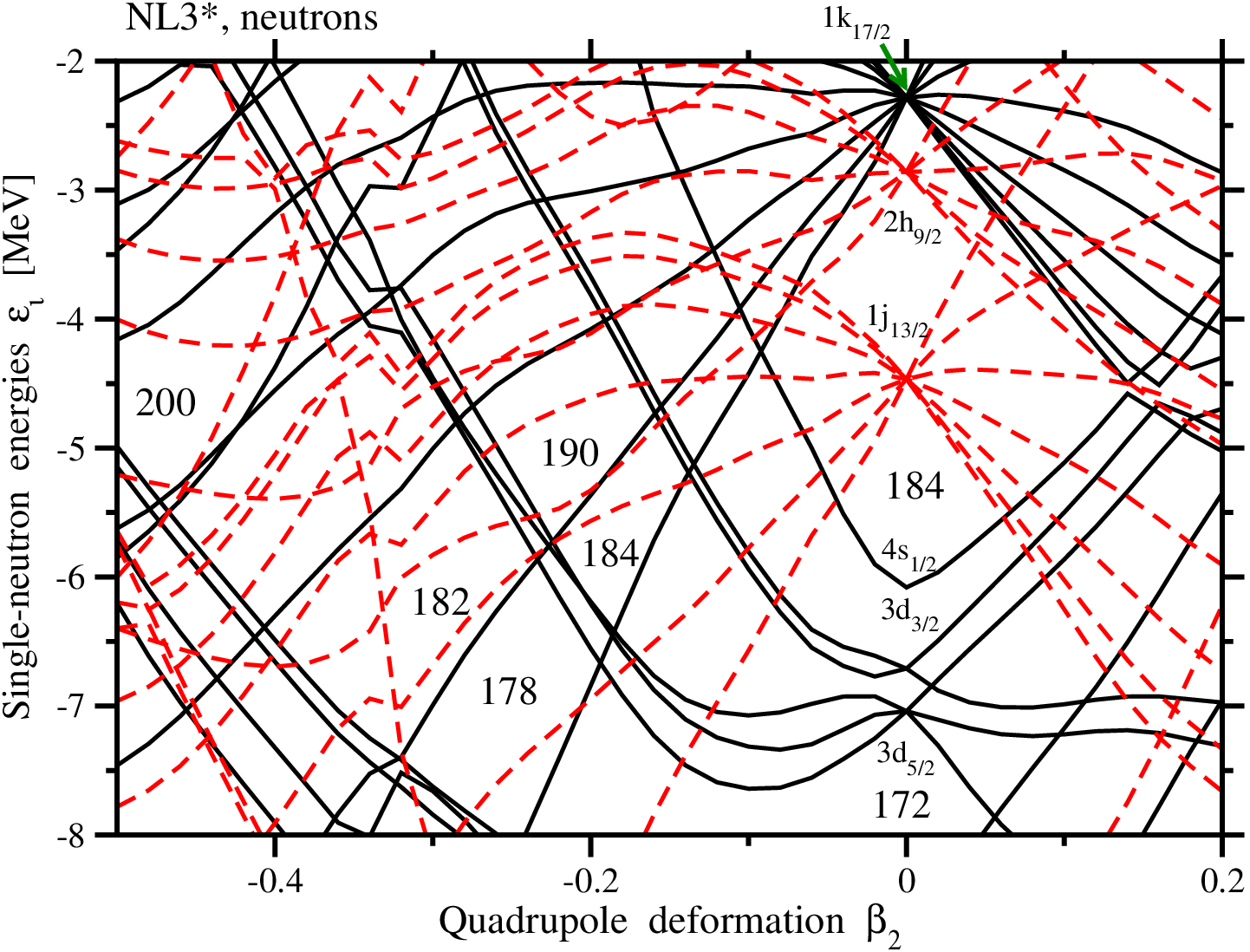}
\caption{(Color online) Single-particle energies, i.e., the diagonal
elements of the single-particle Hamiltonian $h$ in the canonical
basis \cite{RS.80}, for the lowest in total energy solution
in the nucleus $^{304}$120 calculated as a function of the quadrupole
equilibrium deformation $\beta_2$ for the two indicated functionals.
Solid and dashed lines are used for positive and negative parity
states, respectively. Relevant spherical and deformed gaps are
indicated. Note that the transition from spherical to deformation
shapes removes the $2j+1$ degeneracy of the spherical orbitals.
The selected range in deformation is representative for ground state
deformations of the SHEs in the vicinity of the $Z=120$ and $N=184$ lines
and beyond them.}
\label{def-Nilsson}
\end{figure*}

 This is clearly seen in the Nilsson diagrams presented in
Fig.\ \ref{def-Nilsson} for the nucleus $^{304}$120.
Pronounced deformed shell gaps in the proton subsystem are
clearly seen for $\beta \sim -0.2$ at $Z=116,126$,
for $\beta \sim -0.3$ at $Z=118,120$, and for $\beta \sim -0.4$
at $Z=118,122$. Although, in detail, the size of these deformed
gaps, some of which are comparable in magnitude with the $Z=120$
spherical gap, depends on the functional, they are
present both for DD-PC1 and for NL3*. The most pronounced deformed
neutron shell gap is seen at $N=178$ for $\beta_2\sim -0.25$; the
size of this gap is comparable with the spherical shell gap at $N=184$.
At similar deformations somewhat smaller deformed shell gaps
are seen at $N=184, 190, 192$.

 It is important to recognize that contrary to the spherical
states with a degeneracy of $2j+1$, deformed
states are only two-fold degenerate. This will also impact
the shell correction energy since it depends on the averaged
density of the single-particle states in the vicinity of the Fermi
surface~\cite{Strutinsky1966_YadFiz3-614,Strutinsky1967_NPA95-420,Strutinsky1968_NPA122-1}.
As a result, close to the above discussed deformed shell gaps
the negative shell correction energy can be larger in absolute
value than the one at spherical shape even for similar sizes
of the respective deformed and spherical shell gaps.
In the language of the microscopic+macroscopic approach, this difference can be sufficient
to counteract the increase of the  energy of the liquid drop
with increasing oblate deformation in SHEs \cite{NilRag-book}.
The consequences of this interplay between shell correction and
liquid drop energies and the role played by the low level density
of the single-particle states in the vicinity of above discussed
deformed shell gaps are clearly visible in the potential energy curves
of the $^{304}$120 nucleus presented in Fig.\ \ref{Z=120-pes} for
DD-PC1 and NL3*. For DD-PC1, the ground state is oblate with
deformation $\beta_2 \sim -0.3$.  However, two excited minima are
also seen at $\beta_2 \sim -0.15$ and $\beta_2=0.0$.
Although the ground state of the nucleus $^{304}$120 is spherical
for NL3*, three minima at $\beta_2 \sim -0.4$,
$\beta_2 \sim -0.3$, and $\beta_2 \sim -0.2$ are seen at
excitation energies of around 1 MeV. These local minima are
the consequence of the fact that the corresponding minima in the proton
and neutron shell correction energies correspond to different
deformations.

%%%%%%%%%%%%%%%%%%%%%%%%%%%%%%%%%%%%%%%%%%%%%%%%%%%%%%%%%%%%%%
%%%%%%%%%%%%%%%%%%%%%%%%%%%%%%%%%%%%%%%%%%%%%%%%%%%%%%%%%%%%%%%%%%

Similar features are also observed for the $N=184$ isotones
in Fig.\ \ref{N=184-pes}. The nucleus Hs ($Z=108$) has a well pronounced
spherical minimum for NL3*, DD-ME2, and PC-PK1. For these
functionals, the increase of proton number $Z$ leads to an increase of
softness in the potential energy surface for $-0.4 \leq \beta_2 \leq 0$.
However, the ground state remains spherical up
to $Z=120$. On the other side, the ground state of the $Z=124$ nucleus
becomes oblate in these three functionals. The situation is completely
different for the functionals DD-ME$\delta$ and DD-PC1 for which  all
nuclei shown in Fig.\ \ref{N=184-pes} are characterized by soft potential
energy curves in the range $-0.4 \leq \beta_2 \leq 0$ and by
oblate ground states.

%%%%%%%%%%%%%%%%%%%%%%%%%%%%%%%%%%%%%%%%%%%%%%%%%%%%%%%%%%%%%%
\subsection{Comment on superdeformed minima and outer fission barriers}
\label{Out_fission}
%%%%%%%%%%%%%%%%%%%%%%%%%%%%%%%%%%%%%%%%%%%%%%%%%%%%%%%%%%%%%%

The axial RHB calculations restricted to reflection symmetric shapes show
that there exists a second (superdeformed [SD]) minimum with deformation of $\beta_2 \sim 0.5$
or higher for all the nuclei under investigation (Figs.\ \ref{Z=120-pes} and
\ref{N=184-pes}). In the nucleus $Z=120,N=172$ it is in energy close to the
ground state for NL3* and DD-ME2 but lower in energy than the spherical
or oblate minimum for DD-ME$\delta$, DD-PC1 and PC-PK1 (see Fig.\
\ref{Z=120-pes}). With the increase of neutron number the SD minimum becomes
the lowest in energy in all nuclei (see also Ref. \cite{REN-ZZ2001_NPA689-691}).
A similar situation is also observed for the $N=184$ isotones. In the $Z=108$ and $Z=112$
isotopes, the spherical minimum is the lowest in energy for NL3*, DD-ME2 and PC-PK1 (see  Fig.\
\ref{N=184-pes}). With increasing proton number $Z$, the superdeformed minimum
becomes the lowest in energy. The situation is different for DD-ME$\delta$ and
DD-PC1 because (i) in these functionals the superdeformed minimum is the lowest
in energy in all nuclei shown in Fig.\ \ref{N=184-pes}
and (ii) the potential energy curves of the $Z=108, 112$ and 116 nuclei are
much softer [with relatively small inner and outer fission barriers] than in other
functionals. It is necessary to conclude that the relative energies of the
spherical/oblate and superdeformed minima strongly depend on the functional.

Whether these superdeformed states are stable, metastable, or unstable should be
defined by the height (with respect of the SD minimum) and the width of
the outer fission barrier. The results presented in Figs.\ \ref{Z=120-pes}
and \ref{N=184-pes}
show that
while in some nuclei this barrier is appreciable, it is extremely small in
others. Moreover, it was demonstrated in systematic RMF+BCS
calculations with NL3*  for the $Z=112-120$ nuclei that the inclusion
of triaxial or octupole deformation decreases this barrier substantially
by 2 to 4 MeV so it is around or less than  2 MeV in the nuclei studied
in Ref.\ \cite{AAR.12}. Calculations with DD-PC1 and DD-ME2 for
six nuclei centered around $Z=114, N=176$ led to similar results
\cite{AAR.12}.

The impact of octupole deformation on the
outer fission barriers of SHEs has
also been studied in the RMF+BCS calculations with the CEDFs NL3 and NL-Z2 in
Ref.\ \cite{BBM.04}. In this work, the SD minima exist in the calculations
without octupole deformation (see Fig.\ 5 in Ref.\ \cite{BBM.04}). Since the
heights of outer fission barriers in the axial reflection symmetric
calculations is lower in NL-Z2 than in NL3, the inclusion of octupole
deformation completely eliminates the outer fission barriers in NL-Z2 but keeps
their heights around 2.5 MeV in the NL3 functional. In addition, it was
demonstrated in actinides in the RMF+BCS calculations of Ref.\ \cite{LZZ.14} that
non-axial octupole deformation can further reduce the height of outer fission
barrier by $0.5-1$ MeV. Similar effect may be expected also in superheavy nuclei.

  Thus, a detailed analysis of the outer fission barriers requires symmetry
unrestricted calculations in the RHB framework which are extremely time-consuming.
So far no such studies exist. Even the symmetry unrestricted RMF+BCS calculations
(which are at least by one order of magnitude less time consuming than the RHB
calculations) have been performed only for actinides \cite{LZZ.12,LZZ.14,ZLVZZ.15}
in which experimental data on outer fission barriers exist. Global symmetry
unrestricted RHB calculations for SHE have to be left for the future. On  the other
hand, from the discussion above it is clear that outer fission barriers are
expected to be around 2 MeV or less. The results of Refs.\ \cite{BBM.04,AAR.12}
do not cover SHE beyond $Z=120$ and $N=184$ lines. However, Figs.\ \ref{Z=120-pes}
and \ref{N=184-pes} show a clear trend for the decrease of the height of outer
fission barrier beyond these lines; for these nuclei its height is less than 2
MeV in many functionals even in axially symmetric calculations restricted to
reflection symmetric shapes. Thus, it is already clear that the low outer fission
barriers with barrier heights around 2 MeV or less existing in the majority of the
CDFT calculations discussed above would translate into a high penetration probability
for spontaneous fission, such that most likely these superdeformed states (even if
they exist) are metastable. Moreover, non-relativistic calculations usually do not
produce a superdeformed minimum and an outer fission barrier in superheavy nuclei
\cite{BBM.04,KJS.10,MSI.09,SBN.13,WE.12}.

In addition, existing experimental data on SHE (such as total evaporation-residue
cross section or spontaneous fission half-lives) \cite{OU.15} do not show any abrupt
deviation from the expected trends which could be interpreted as a transition to a
superdeformed ground state.

 These are the reasons why we restrict our consideration to the ground states
associated with either normal-deformed prolate, oblate or spherical
minima.
% and consider in the following only the inner fission barrier.

%%%%%%%%%%%%%%%%%%%%%%%%%%%%%%%%%%%%%%%%%%%%%%%%%%%%%%%%%%%%%%%%%
\section{The systematics of the deformations}
\label{def-system}
%%%%%%%%%%%%%%%%%%%%%%%%%%%%%%%%%%%%%%%%%%%%%%%%%%%%%%%%%%%%%%%%%

\begin{figure*}[ht]
\centering
\includegraphics[angle=-90,width=8.8cm]{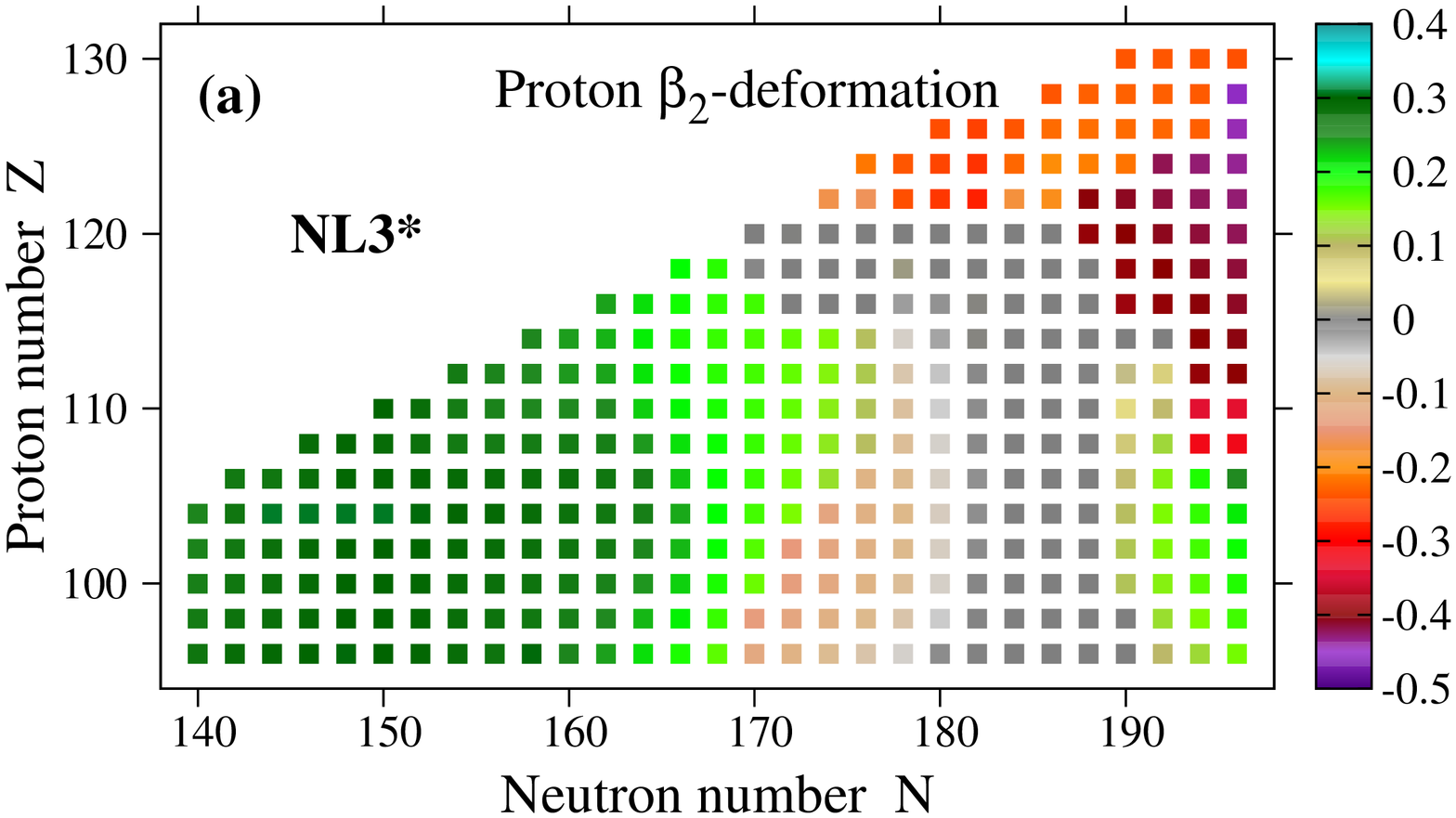}
\includegraphics[angle=-90,width=8.8cm]{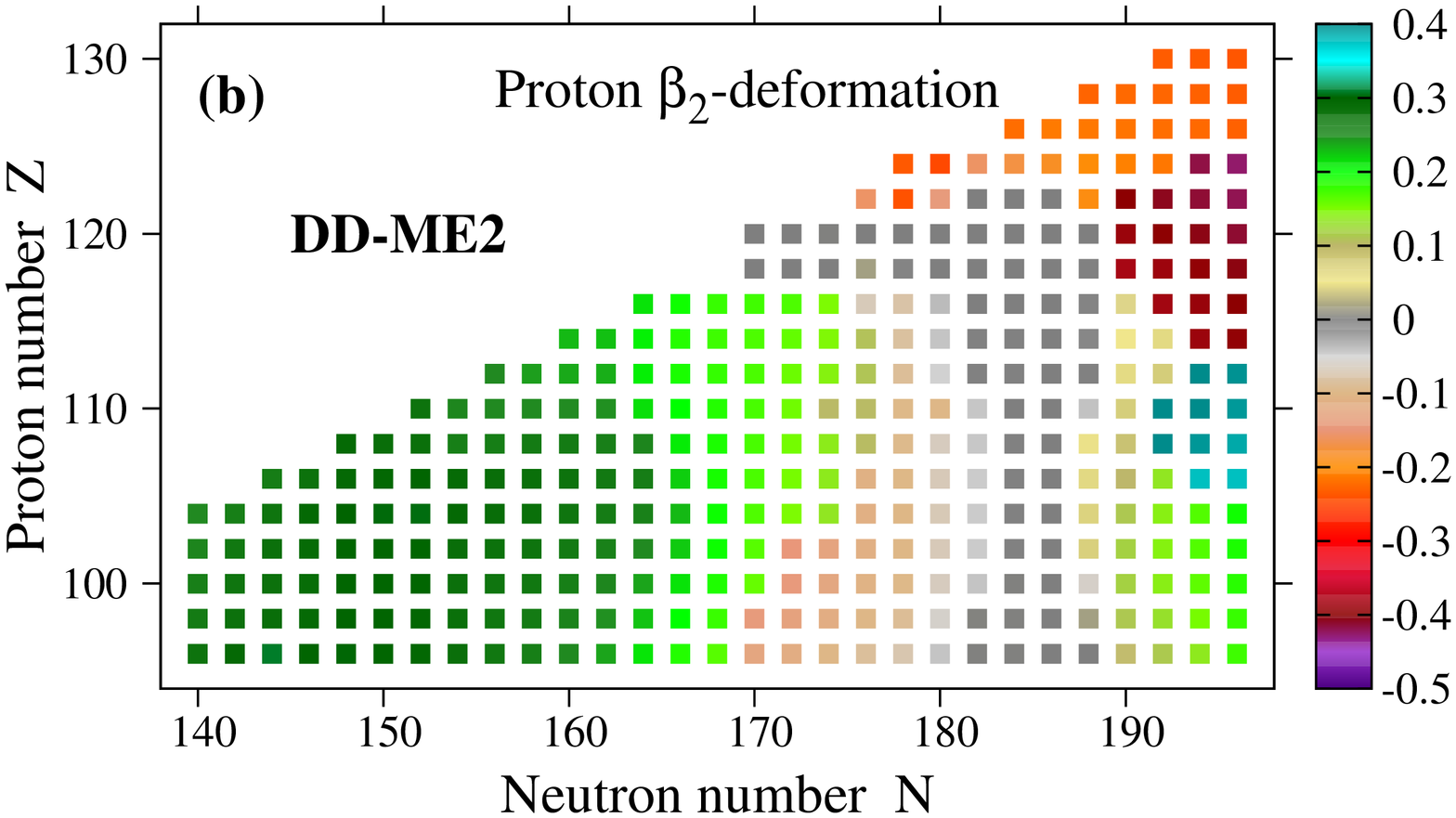}
\includegraphics[angle=-90,width=8.8cm]{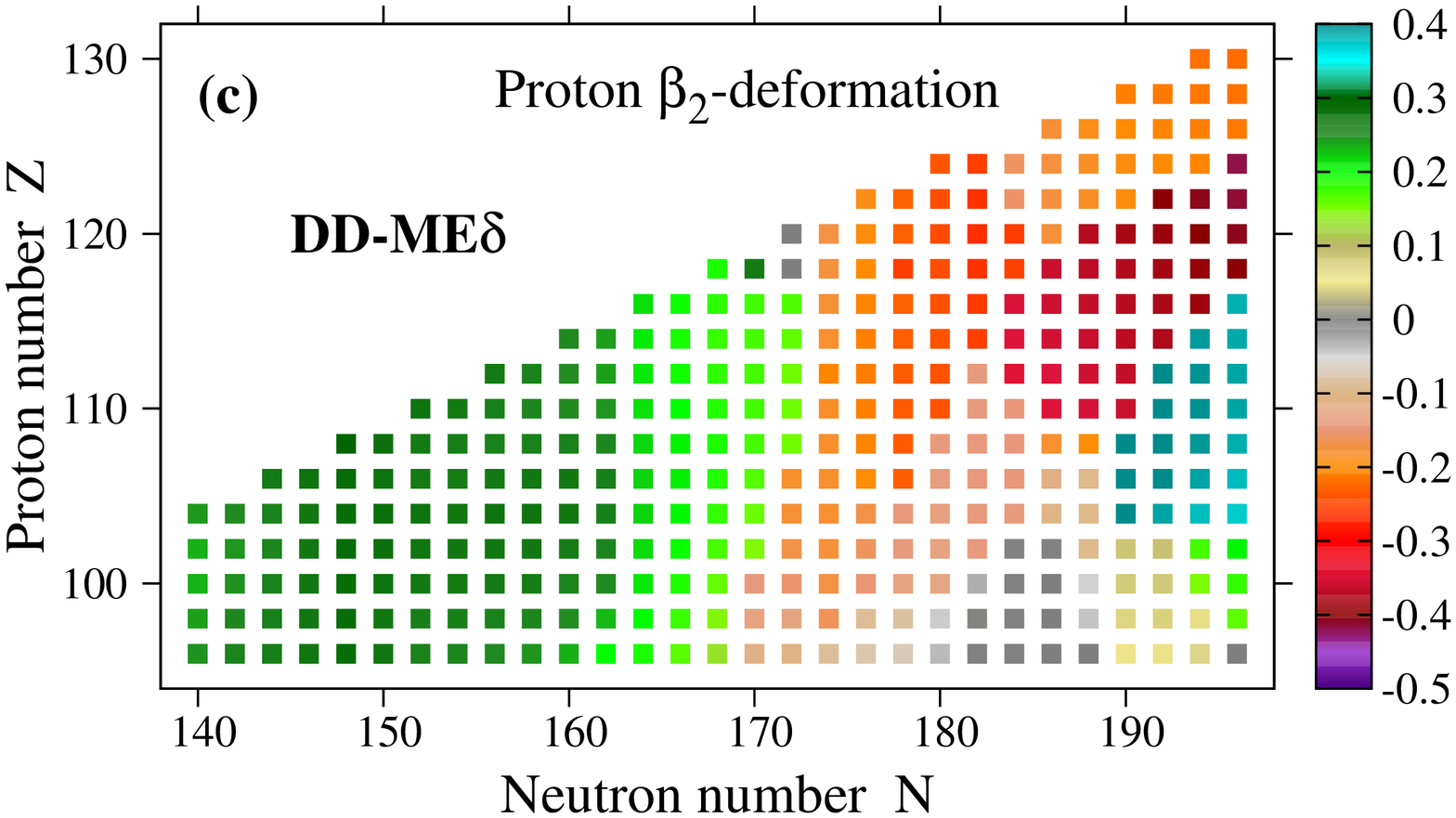}
\includegraphics[angle=-90,width=8.8cm]{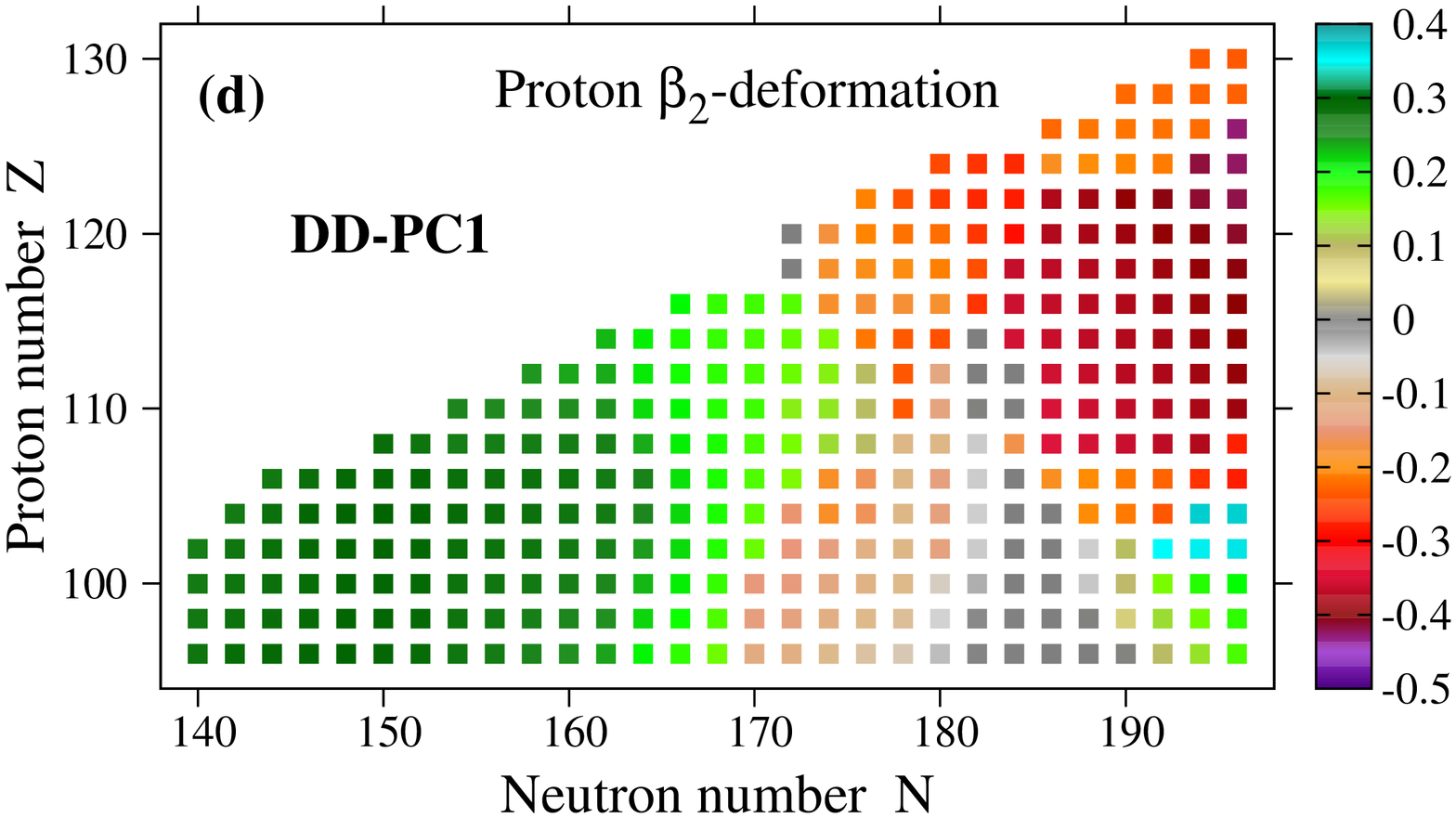}
\includegraphics[angle=-90,width=8.8cm]{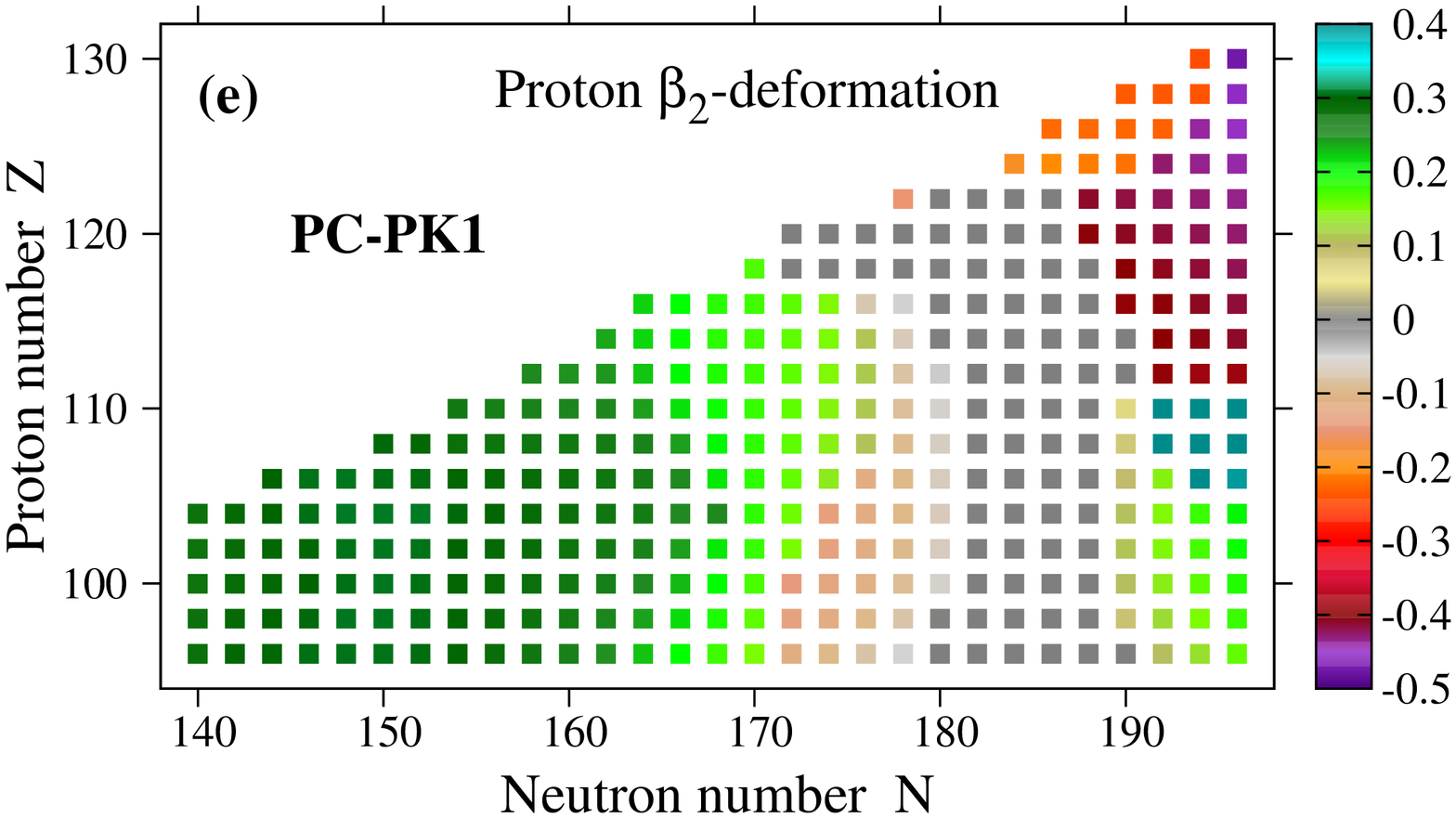}
\caption{(Color online) Charge quadrupole deformations
$\beta_2$ obtained in the RHB calculations with the indicated CEDFs
\protect\footnote{Note
that ground state equilibrium deformations of the $N=182$ nuclei
with $Z=106$ and 116 and of the $N=184$ nuclei with $Z=108, 114, 116$
and $118$ obtained with DD-PC1 (Fig.\ \ref{deformation}d)
differ from the ones shown in the bottom panel of Fig.\ 4 in Ref.\
\protect\cite{AA.15}. This is a consequence of the use of a smaller
deformation range ($-0.4\leq\beta_2\leq 1.0$) and a larger
step in deformation ($\Delta \beta_2=0.05$) in the RHB calculations of
Ref.\ \protect\cite{AA.15} as compared with the present manuscript.}.
}
\label{deformation}
\end{figure*}

The calculated charge quadrupole deformations of the ground states
for five CEDFs are plotted in Fig.\ \ref{deformation}. They are shown
for the $Z=96-130$ nuclei located between the two-proton drip line (see Table IV in
Ref.\ \cite{AARR.14}) and $N=196$. The impact of the large spherical shell
gaps discussed in Sect.\ \ref{sp-states} on the structure of superheavy
nuclei can be accessed via the analysis of the width of the band of
spherical nuclei shown by gray color in the $(Z,N)$ chart of Fig.\
\ref{deformation}. The width of this gray region along a specific
particle number corresponding to a shell closure indicates the impact
of this shell closure on the structure of neighbouring nuclei. For example,
for NL3* the width of such a band at $Z\approx 120$ is on average three even-even nuclei
in the $Z$ direction for $N=172-188$ and the width of a corresponding band
at $N\approx 184$ is on average four even-even nuclei in the
$N$ direction for $Z=96-120$. This result is contrary to existing discussions
in covariant density functional theory which emphasize the impact of the $N=172$
shell gap over the $N=184$ gap. Our results clearly show that the effect of the $N=184$ spherical
shell gap on the equilibrium deformation is more pronounced as compared with
the $N=172$ gap. A similar situation exist in the calculations with PC-PK1
(see Fig.\ \ref{deformation}e). However, the impact of the $Z=120$
and the $N=184$ spherical shell gaps becomes less pronounced for DD-ME2
(see Fig.\ \ref{deformation}b).

The impact of the $Z=120$ spherical shell gap is significantly reduced
for DD-ME$\delta$ and DD-PC1; only the $N=172$
nuclei with $Z=118$ and $120$ are spherical for those two functionals.  The
impact of the $N=184$ shell gap is also considerably decreased; the ground
states of the $N=184$ nuclei are spherical only for $Z\leq  102$ in DD-ME$\delta$
and for $Z\leq 112$  in DD-PC1 (see Figs.\ \ref{deformation}c and d).
Note also that the band of spherical nuclei around $N=184$ is narrow for DD-PC1.
These results are in contradiction to the expectation that
the large size of the spherical $Z=120$ gap in Fig.\ \ref{spectra} forces
the isotopes with $Z=120$ to be spherical for a large range of neutron numbers.
Note that proton and neutron shell gaps act simultaneously in the vicinity of
a nucleus with proton and neutron numbers corresponding to those gaps.
Thus, the effect of a single gap is more quantifiable away from this nucleus.

It is interesting to compare these results with the ones obtained
for $Z\leq 104$ nuclei in Fig. 17 of Ref.\ \cite{AARR.14} with NL3*,
DD-ME2, DD-ME$\delta$ and DD-PC1. In these nuclei the neutron
$N = 82, 126$, and 184 shell gaps have a more pronounced effect on
the nuclear deformations as compared with the proton shell gaps at
$Z = 50$ and $Z = 82$. This feature was common to all the CEDFs
used in Ref.\ \cite{AARR.14}. However, the width of the band of spherical
and near-spherical nuclei along these neutron numbers was broader in
NL3* as compared with other functionals under consideration. We see the same
feature also in superheavy nuclei along the $N=184$ shell closure (see
Fig.\ \ref{deformation}); note that PC-PK1 was not used in Ref.\
\cite{AARR.14}.

The RHB results for superheavy nuclei show the unusual feature that the ground states
of the nuclei outside the band of spherical or near-spherical shapes
(shown by gray color in Fig.\ \ref{deformation}) have oblate shapes for
NL3*, DD-ME2 and PC-PK1. This is contrary to the
usual situation observed in the lighter nuclei (see, for example,
Fig. 17 in Ref.\ \cite{AARR.14}) where, between shell
closures, the nuclei change their shape with increasing particle numbers from spherical to
prolate, then to oblate, and finally back to spherical.
It is interesting to see that the systematic microscopic+macroscopic (MM) calculations
of Ref.\ \cite{JKS.11} based on the Woods-Saxon potential also show a
similar preponderance of oblate shapes in the ground states of superheavy nuclei.
These MM results were also confirmed by the calculations for a few nuclei performed within
Skyrme DFT with the functional SLy6 \cite{JKS.11}. The situation is even more
drastic for the CEDFs DD-ME$\delta$ (DD-PC1) in which no (or very limited)
indications of spherical shapes are seen on passing through the nuclei with
$Z=120$ or $N=184$. When comparing our results with other calculations one has
to keep in mind that not all published results extend to a sufficiently large
deformation for oblate shapes (see, for example, Ref.\ \cite{WE.12}).

%%%%%%%%%%%%%%%%%%%%%%%%%%%%%%%%%%%%%%%%%%%%%%%%%%%%%%%%%%%%%%
%%%%%%%%%%%%%%%%%%%%%%%%%%%%%%%%%%%%%%%%%%%%%%%%%%%%%%%%%%%%

\begin{figure*}[ht]
\centering
\includegraphics[angle=-90,width=8.8cm]{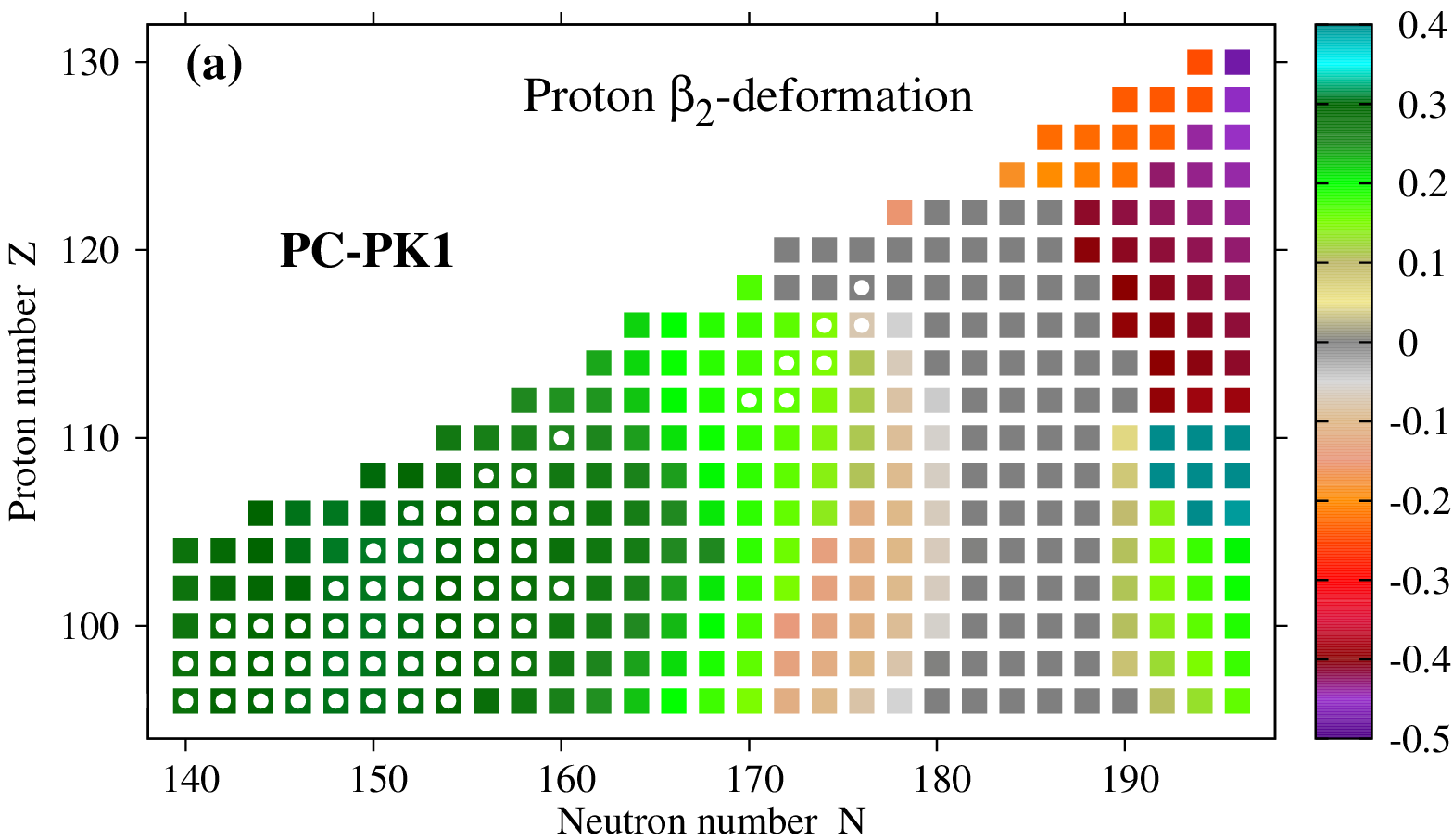}
\includegraphics[angle=-90,width=8.8cm]{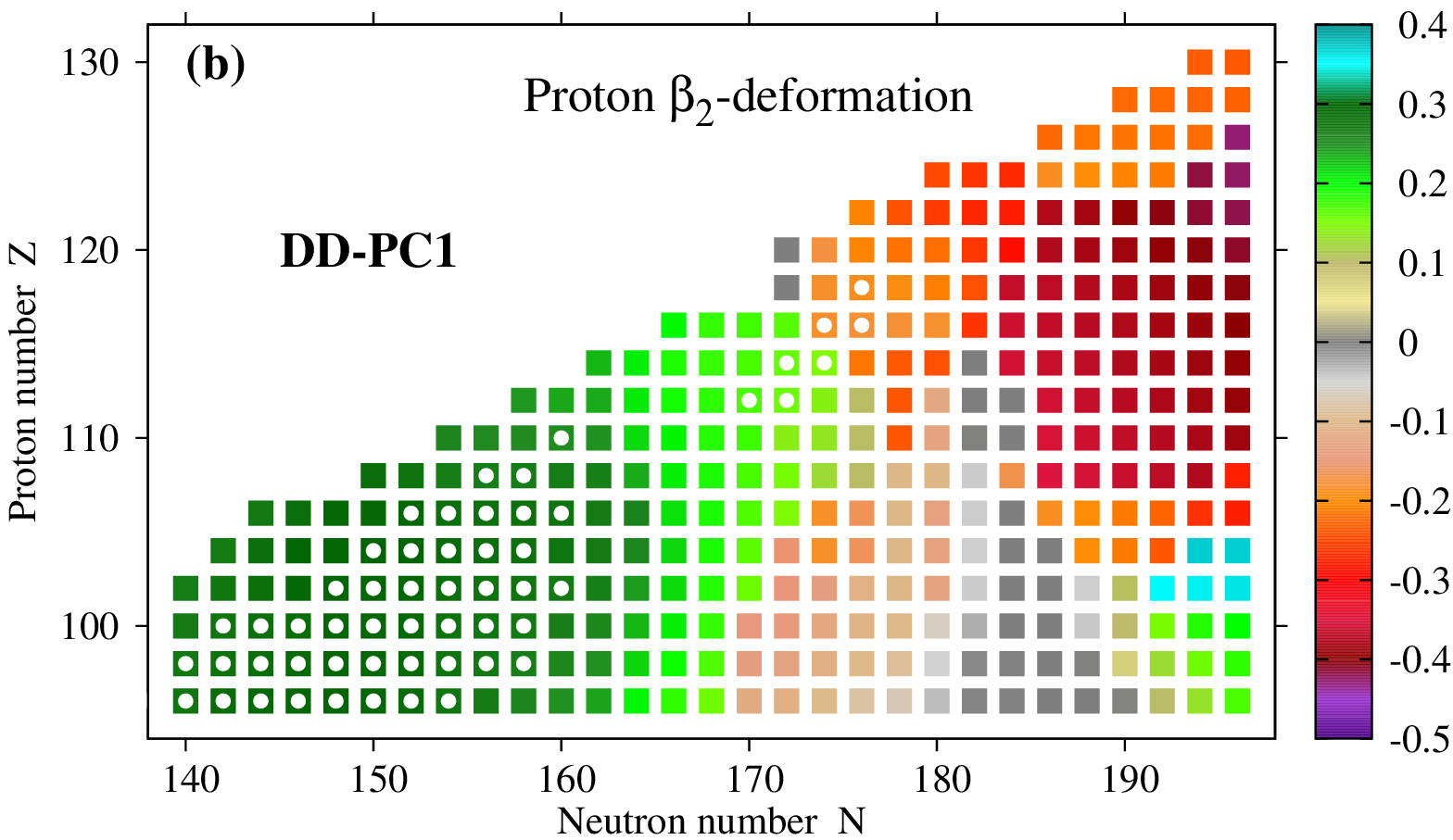}
\caption{(Color online) The same as Fig.\ \ref{deformation} but
with experimentally known nuclei shown by open circles. Only the
results with PC-PK1 and DD-PC1 are shown. The information on 
experimentally known nuclei is taken from Refs.\ \cite{OU.15,Eval-data}. 
Note that the region of experimentally known superheavy nuclei is 
broader at high $Z$ when odd and odd-odd mass nuclei are included 
(see Fig.\ 2 in Ref.\ \cite{OU.15}).}
\label{defor-known-nuclei}
\end{figure*}

 Fig.\ \ref{defor-known-nuclei} shows the map of calculated
charge quadrupole deformations with experimentally known nuclei
indicated by open circles. One can see that, apart from the
$Z=116, 118$ nuclei, the predictions of these two functionals
(PC-PK1 and DD-PC1)
for the equilibrium deformations of experimentally known
even-even nuclei are very similar. For these nuclei,
PC-PK1 predicts the gradual transition from prolate to
spherical shape on going from $Z=114$ to $Z=118$. The same
happens also for NL3* and DD-ME2 (Fig.\
\ref{deformation} a and b). On the contrary, for DD-PC1 the transition
from the prolate to oblate minimum is predicted for experimentally
known nuclei on going from $Z=114$ to
$Z=116$ and all experimentally known $Z\geq 116$ nuclei are
expected to be oblate. The same happens also for
DD-ME$\delta$. However, because of the limited scope
of experimental data these differences in the description of
experimentally known $Z=116$ and $118$ nuclei between
DD-PC1/DD-ME$\delta$ and PC-PK1/NL3*/DD-ME2 cannot
be discriminated.

%%%%%%%%%%%%%%%%%%%%%%%%%%%%%%%%%%%%%%%%%%%%%%%%%%%%%%%%%%%%%%
%%%%%%%%%%%%%%%%%%%%%%%%%%%%%%%%%%%%%%%%%%%%%%%%%%%%%%%%%%%%

\begin{figure*}[ht]
\centering
\includegraphics[angle=-90,width=11cm]{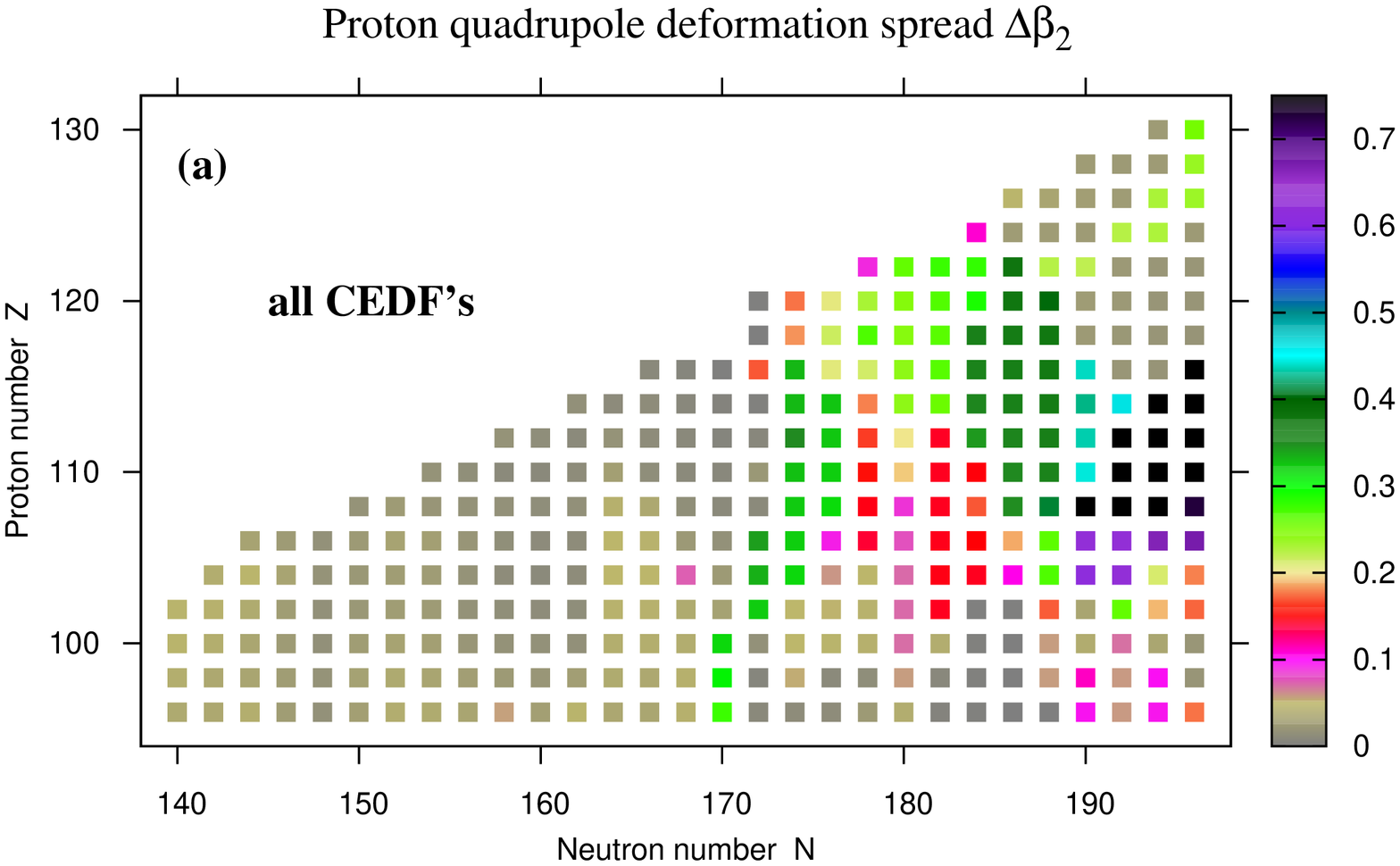}
\includegraphics[angle=-90,width=11cm]{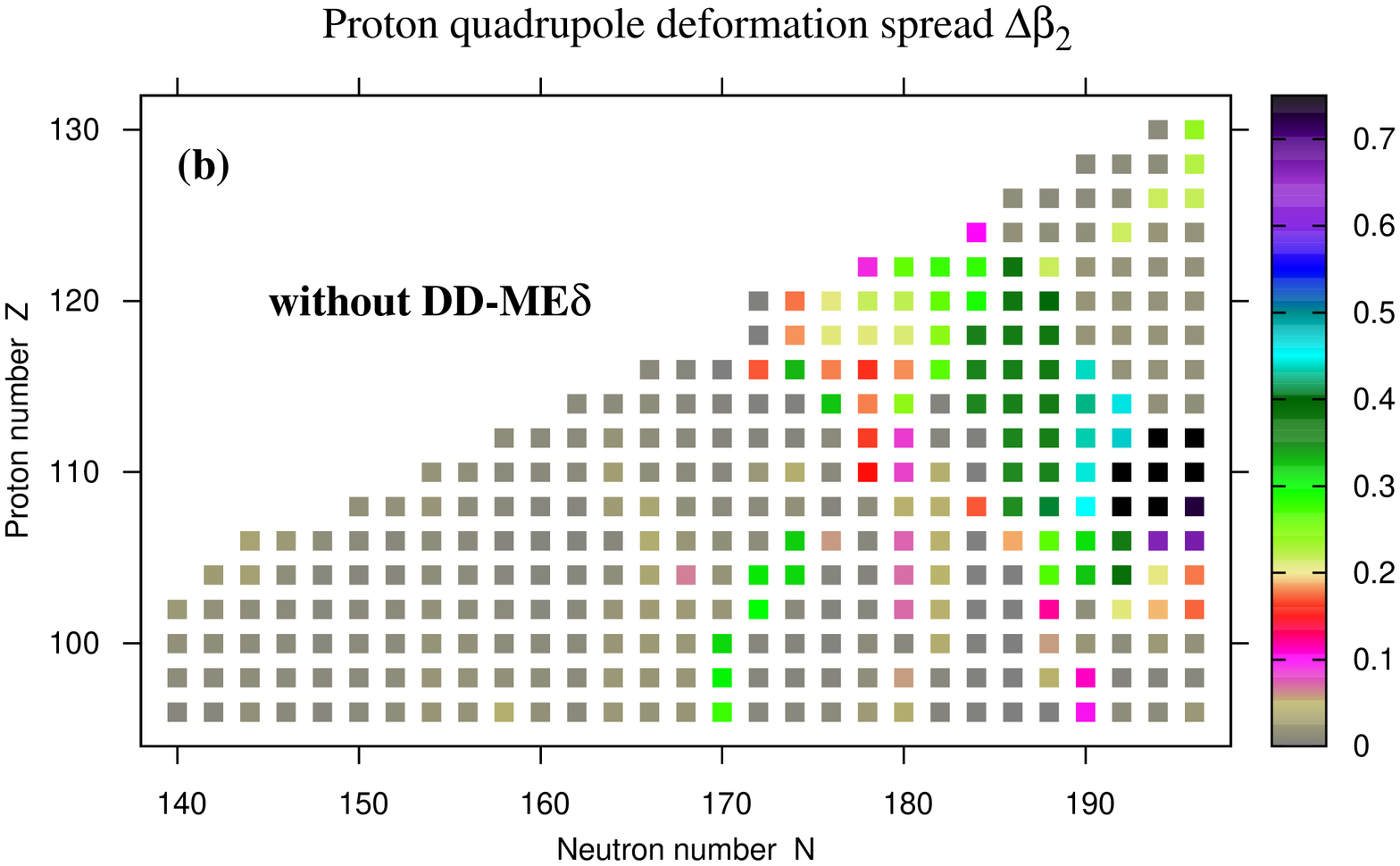}
\caption{(Color online) Proton quadrupole deformation spreads
$\Delta \beta_2$ as a function of proton and neutron number.
$\Delta \beta_2(Z,N) = |\beta_2^{max}(Z,N) - \beta_2^{min}(Z,N)|$, where
$\beta_2^{max}(Z,N)$ and $\beta_2^{min}(Z,N)$ are the largest and
smallest proton quadrupole deformations obtained with the set
of CEDFs used for the nucleus $(Z,N)$. Panel (a) shows  the results for
all functionals, while DD-ME$\delta$ is excluded in the results shown on panel (b).}
\label{deformation_spread}
\end{figure*}

 Spreads in the theoretical predictions of the charge quadrupole
deformations are shown in left panel of Fig.\ \ref{deformation_spread}.
They are very small in the region of known nuclei and for $N<170$. Only
very few experimentally known nuclei with  $Z=114, 116$ and 118 are
located in the region where substantial theoretical spreads exist
(compare Fig.\ \ref{deformation_spread} with Fig.\ \ref{defor-known-nuclei}).
However, as discussed above available experimental data on these nuclei
does not allow to discriminate different predictions. Quite large
spreads exist in the region near the $Z=120$ and $N=184$
lines. This is because spherical ground states are predicted in this
region by NL3*, DD-ME2 and PC-PK1, while DD-ME$\delta$ and DD-PC1
favor oblate shapes in these nuclei. Very large spreads exist
in the $Z\sim 110, N\geq 190$ region; this is a region where a transition
from prolate to oblate shape is seen in the calculations and it
takes place at different positions in the $(Z,N)$ chart for the different
functionals (see Fig.\ \ref{deformation}). The theoretical spreads
become small again in the upper right corner of the chart; here they are
substantial only in several nuclei (shown by green color) which form a ``line'' parallel to the two-proton
drip line. This ``line'' is a consequence of the fact that the transition from
ground state deformations $\beta_2 \sim -0.2$ to $\beta_2 \sim -0.4$
takes place for different functionals at different positions in the $(N,Z)$
chart (see Fig.\ \ref{deformation}).

The right panel of Fig.\ \ref{deformation_spread} shows theoretical
spreads for the case when the functional DD-ME$\delta$ is excluded
from consideration. This functional provides unrealistically low
heights of inner fission barriers in SHEs \cite{AANR.16}
% (see discussion in Sec.\
% \ref{syst-fission-barrier} and conclusions)
and thus  it is very unlikely that this functional is appropriate 
for the region of SHEs. However, its exclusion from consideration 
reduces only slightly the theoretical spreads.

It is interesting to compare the results of the present analysis
of theoretical spreads in the description of ground state
deformations with the global analysis presented in Ref.\ \cite{AARR.14}
for $Z\leq 104$ nuclei. It is clear that the region of SHEs in the
vicinity of the $Z=120$ and $N=184$ lines bears the mark of a
transitional region characterized either by soft potential energy
surfaces or by shape coexistence. This is the source of large theoretical
spreads in the prediction of ground state deformations
which exist not only in the region of SHEs but also globally
(see Ref.\ \cite{AARR.14}). Both in SHEs and globally these
uncertainties are attributable to the deficiencies of the
current generation of functionals with respect to the
description of single-particle energies.

%%%%%%%%%%%%%%%%%%%%%%%%%%%%%%%%%%%%%%%%%%%%%%%%%%%%%%%%%%%%%%
%%%%%%%%%%%%%%%%%%%%%%%%%%%%%%%%%%%%%%%%%%%%%%%%%%%%%%%%%%%%

%%%%%%%%%%%%%%%%%%%%%%%%%%%%%%%%%%%%%%%%%%%%%%%%%%%%%%%%%%%%%%%%%%%%%%%%%%%%%%%%%
\section{The quantities $\delta_{2n}(Z,N)$ and $\delta_{2p}(Z,N)$ as
indicators of shell gaps}
\label{Sect-delta_2}
%%%%%%%%%%%%%%%%%%%%%%%%%%%%%%%%%%%%%%%%%%%%%%%%%%%%%%%%%%%%%%%%%%%%%%%%%%%%%%%%%

 The analysis of the shell structure (and shell gaps) of superheavy nuclei
is most frequently  based  on the quantity $\delta_{2n}(Z,N)$ defined as (Ref.\
\cite{BRRMG.99,A250})
\begin{eqnarray}
\delta_{2n}(Z,N)=S_{2n}(Z,N)-S_{2n}(Z,N+2)=~~~~~~~~~\\
=-B(Z,N-2)+2B(Z,N)-B(Z,N+2).\nonumber
\label{2n-shell-gap}
\end{eqnarray}
Here $B(Z,N)$ is the binding energy and $S_{2n}(Z,N)$ is the two-neutron separation
energy. The quantity $\delta_{2n}(Z,N)$, being related to the second derivative
of the binding energy as a function of the neutron number, is a more sensitive
indicator of the local decrease in the single-particle density associated with
a shell gap than the two-neutron separation energy $S_{2n}(Z,N)$. This quantity
is frequently called as {\it two-neutron shell gap}. In a similar way, for protons,
$\delta_{2p}(Z,N)$ is defined as
\begin{eqnarray}
\delta_{2p}(Z,N)=S_{2p}(Z,N)-S_{2p}(Z+2,N)=~~~~~~~~~\\
=-B(Z-2,N)+2B(Z,N)-B(Z+2,N).\nonumber
\label{2p-shell-gap}
\end{eqnarray}
%

%%%%%%%%%%%%%%%%%%%%%%%%%%%%%%%%%%%%%%%%%%%%%%%%%%%%%%%%%%%%%%
\begin{figure*}[ht]
\centering
\includegraphics[angle=-90,width=8.8cm]{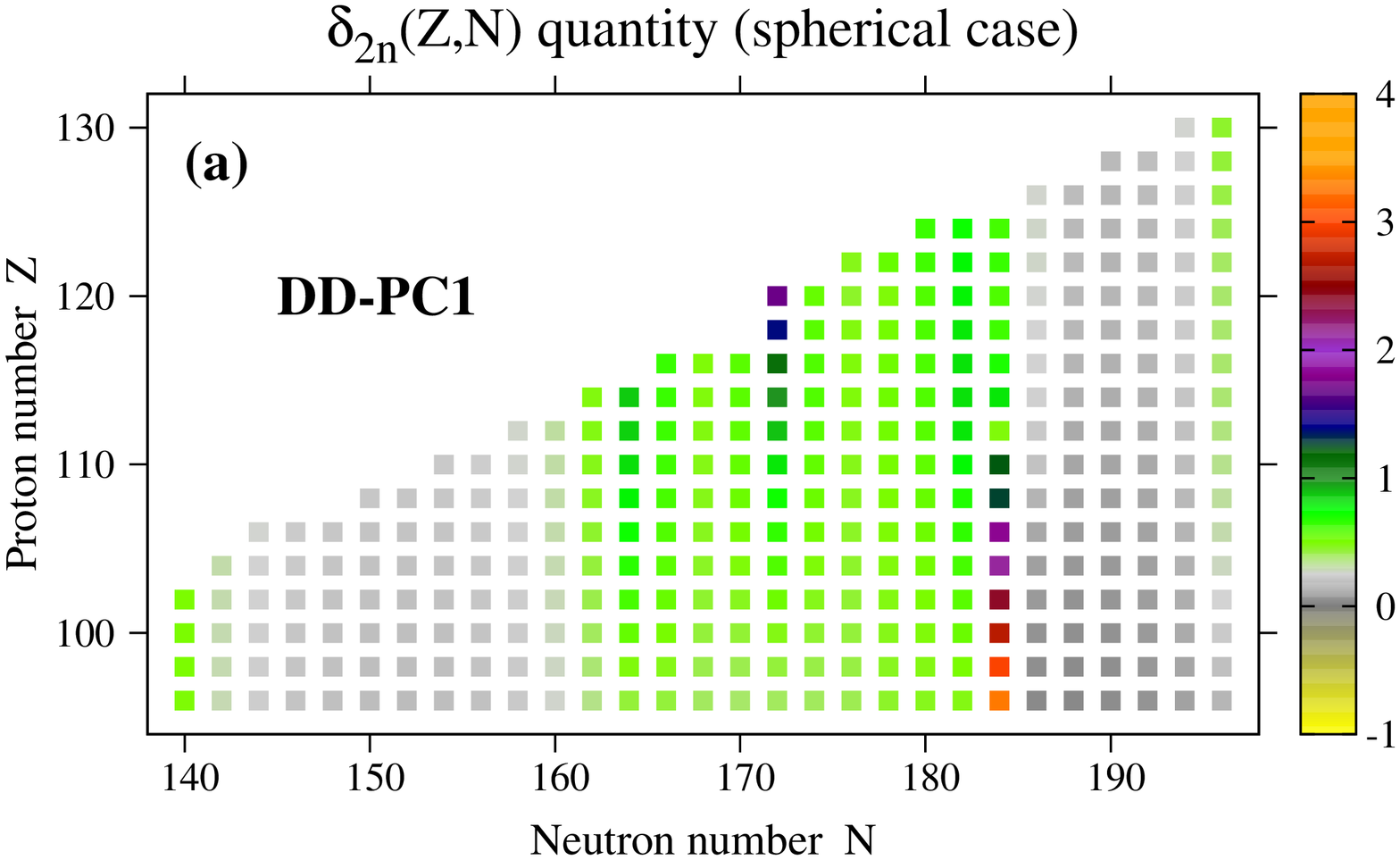}
\includegraphics[angle=-90,width=8.8cm]{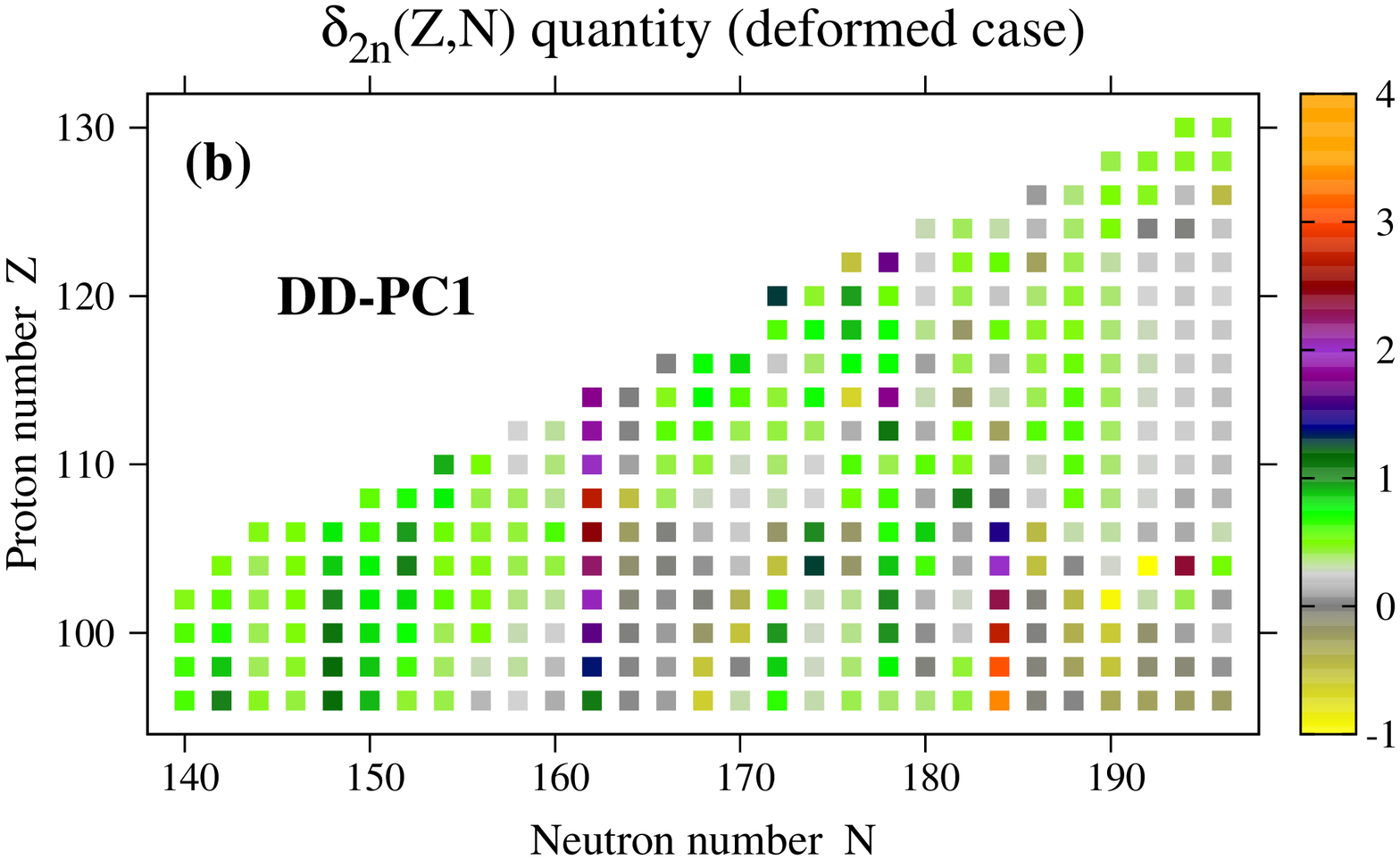}
\includegraphics[angle=-90,width=8.8cm]{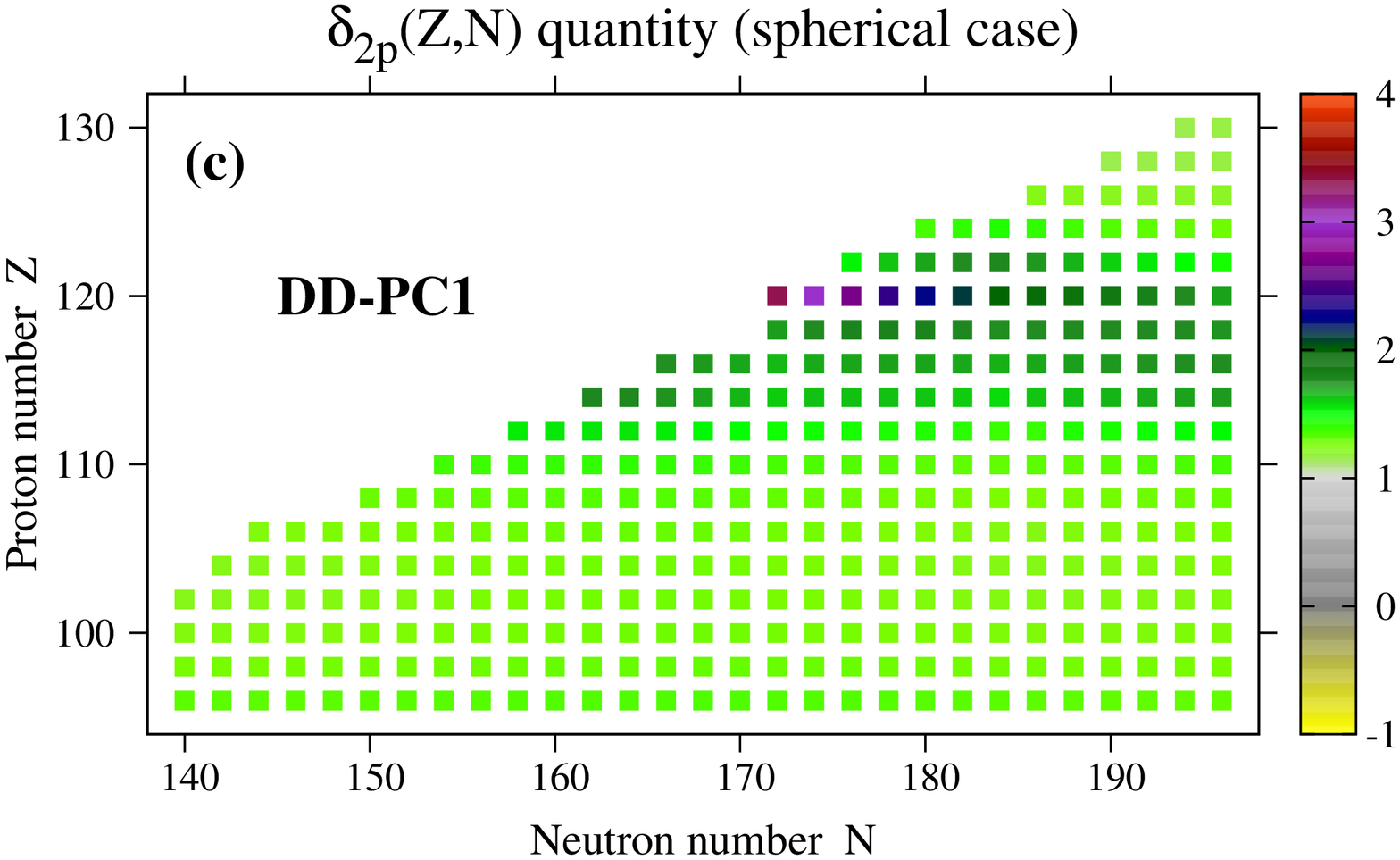}
\includegraphics[angle=-90,width=8.8cm]{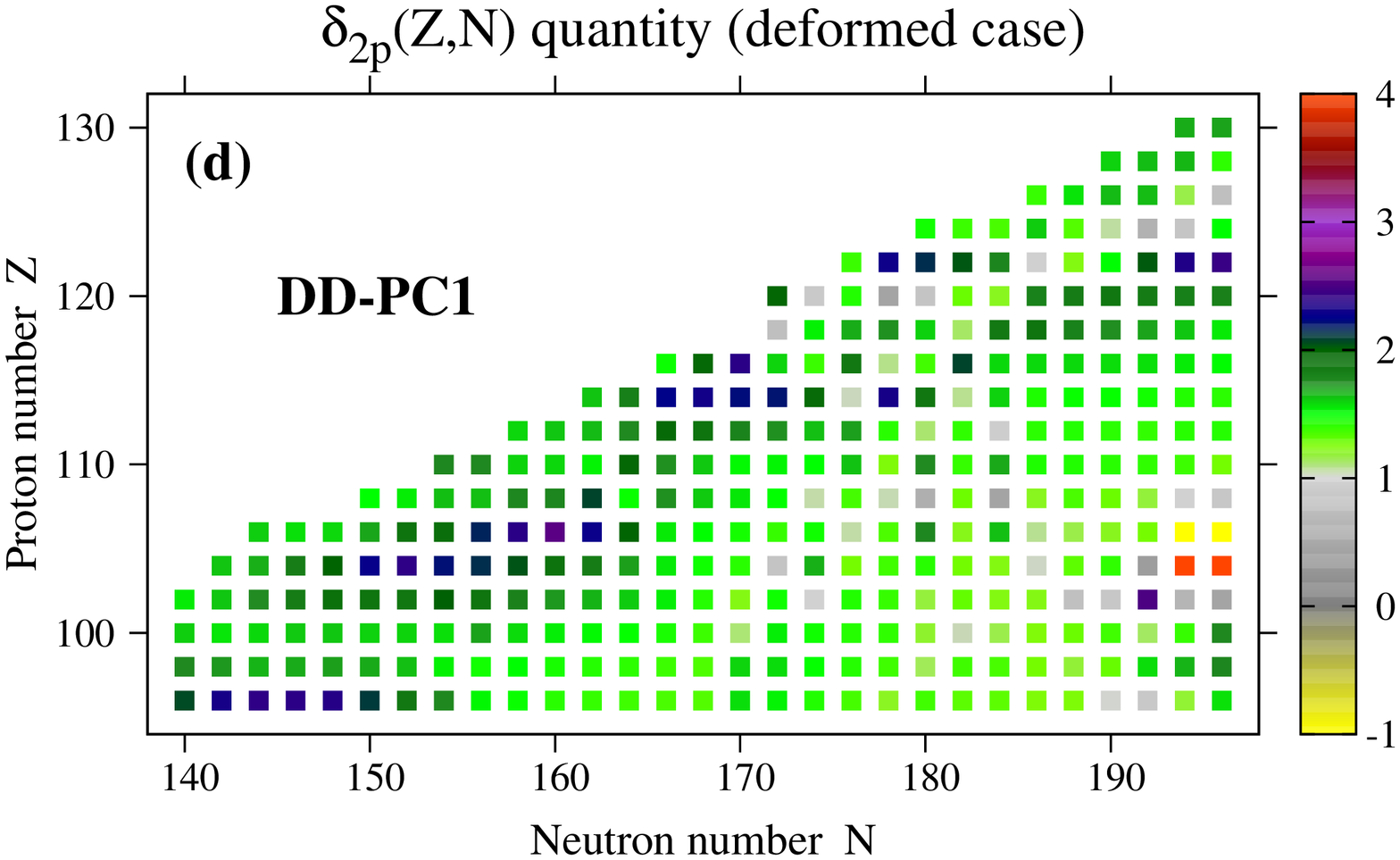}
\caption{(Color online) The comparison of the quantities $\delta_{2n}(Z,N)$
(top panels) and $\delta_{2p}(Z,N)$ (bottom panels) obtained in
spherical (left panels) and deformed (right panels) RHB calculations with
DD-PC1. Note that for clarity of the presentation, the colormaps for
the quantities $\delta_{2p}(Z,N)$ for protons and $\delta_{2n}(Z,N)$
for neutrons are different.}
\label{delta_2n2p_sph_def}
\end{figure*}
%%%%%%%%%%%%%%%%%%%%%%%%%%%%%%%%%%%%%%%%%%%%%%%%%%%%%%%%%%%%%

However, as discussed in detail in Ref.\ \cite{A250}, many factors
beyond the size of the single-particle shell gap contribute to $\delta_{2n}(Z,N)$
and $\delta_{2p}(Z,N)$, as for instance
deformation and pairing changes. For example, the global analysis of these quantities in Ref.\
\cite{AARR.15} shows that for some $(Z,N)$ values $\delta_{2n}(Z,N)$ becomes
negative because of deformation
changes. Since by definition the shell gap has to be positive, it is clear that the
quantities $\delta_{2n,2p}(Z,N)$ cannot serve as explicit measures of the size of the
shell gaps.

Unfortunately, in the majority of cases the analysis of these quantities in
superheavy nuclei (and thus the conclusions about the underlying shell structure
and the gaps) is based on the results of spherical calculations (see, for
example, Refs.\ \cite{BRRMG.99,ZMZGT.05,LG.14}). Thus the possibility of a
considerable softness of the potential energy surface leading to a deformed
minimum is ignored from the beginning. An alternative way to analyze the shell
structure is via the microscopic shell correction energy (see, for example,
Ref.\ \cite{BNR.01}). However, such an analysis is also frequently limited to
spherical shapes (Ref.\ \cite{BNR.01}) and, in addition, the comparison with
experiment is less straightforward.

%%%%%%%%%%%%%%%%%%%%%%%%%%%%%%%%%%%%%%%%%%%%%%%%%%%%%%%%%%%%%%%%%%%%%%%
\begin{table*}[htbp]
 \caption{rms-deviations $\Delta E_{\rm rms}$, $\Delta (S_{2n})_{\rm rms}$
($\Delta (S_{2p})_{\rm rms}$), $\Delta(Q_{\alpha})_{rms}$ and $\Delta(\tau_{\alpha})_{rms}$
between calculated and experimental binding energies $E$, two-neutron(-proton)
separation energies $S_{2n}$ ($S_{2p}$), $Q_{\alpha}$ values and $\alpha$-decay
half-lives $\tau_{\alpha}$. The values of physical observables in the columns 2-5
are presented in the following format ``A/B'', where A are the values obtained
from only measured masses and B from measured+estimated masses. Note that
only experimental data on even-even nuclei with $Z\geq 96$ is used here. In the
last column, the deviations are given in terms of orders of magnitude. In each
column, bold style is used to indicate the functional with the best rms-deviation.}
 \begin{center}
 \begin{tabular}
 [c]{|c|c|c|c|c|c|}\hline
CEDF & $\Delta{E}_{rms}$ [MeV] & $\Delta({S}_{2n})_{rms}$ [MeV] & $\Delta(S_{2p})_{rms}$ [MeV] &
$\Delta(Q_{\alpha})_{rms}$ [MeV] & $\Delta(\tau_{\alpha})_{rms}$ [order] \\ \hline
      1        &     2       &    3        &    4       &    5       &   6        \\ \hline
NL3*           & 3.02/3.39   & 0.71/0.68   & 1.33/1.34  & 0.68/0.75  & 2.44       \\
DD-ME2         & 1.39/1.40   & 0.45/0.54   & 0.85/0.90  & 0.51/0.65  & 1.95       \\
DD-ME$\delta$  & 2.52/2.45   & 0.60/0.51   & 0.45/0.48  & 0.39/0.51  & 1.39       \\
DD-PC1      & {\bf 0.59/0.74} & 0.30/0.32  & 0.41/0.42  & 0.36/0.47  & 1.40       \\
PC-PK1         & 2.82/2.63    & {\bf 0.25/0.23} & {\bf 0.36/0.33} & {\bf 0.32/0.38} & {\bf 1.26} \\
\hline
\end{tabular}
\label{deviat}
\end{center}
\end{table*}
%%%%%%%%%%%%%%%%%%%%%%%%%%%%%%%%%%%%%%%%%%%%%%%%%%%%%%%%%%%%%%%%%%%%%%%%

 The danger of a misinterpretation of the structure of superheavy nuclei based
on the analysis of the quantities $\delta_{2n}(Z,N)$ and $\delta_{2p}(Z,N)$ obtained
in spherical calculations is illustrated in Fig. \ \ref{delta_2n2p_sph_def} on
the example of DD-PC1. In spherical calculations,
the  quantity $\delta_{2n}(Z,N)$ has pronounced maxima at $N=184$ for $Z=96-110$ and
less pronounced maxima at $N=172$ for $Z=112-120$. Note that in this functional
the nucleus $(Z=120, N=172)$ is located beyond the two-proton drip line. However,
it is clear that the impact of the $N=172$ shell gap does not propagate far
away from $Z=120$. The quantity $\delta_{2p}(Z,N)$ is enhanced in a broad
region around $Z=116-120$ and has a maximum for $Z=120$ which becomes
especially pronounced approaching $N=172$.

However, in deformed RHB calculations, the quantities $\delta_{2n}(Z,N)$ and $\delta_{2p}(Z,N)$
show a picture in many respects quite different from the one obtained in spherical calculations.
In addition to the  maxima in
the quantity $\delta_{2n}(Z,N)$ at $N=184$ for $Z=96-108$, which are already seen
in spherical calculations, deformed RHB calculations show maxima at $N=162$
(for $Z=96-112$) and at $N=148$ (for $Z=98-102$). The latter gap appears for a number
of covariant functionals instead of the experimentally observed $N=152$ gap (see Refs.\
\cite{DABRS.15} for details). Note that the maxima in $\delta_{2n}(Z,N)$ seen
at $N=172$ in spherical calculations disappear in deformed RHB calculations. In addition,
some isolated peaks in the quantity $\delta_{2n}(Z,N)$ appear across the nuclear chart
of Fig.\ \ref{delta_2n2p_sph_def}d at specific values of $Z$ and $N$. In many cases,
they originate from rapid deformation changes in going from one nucleus to another.

Even more drastic differences are seen when comparing the quantities $\delta_{2p}(Z,N)$
obtained in spherical and in deformed RHB calculations. Any indication
of the $Z=120$ spherical shell gap clearly visible in the spherical case (Fig.\
\ref{delta_2n2p_sph_def}c), disappear in deformed calculations (Fig.\
\ref{delta_2n2p_sph_def}d). This is a consequence of the fact that apart of the
$Z=118,120$ nuclei with $N=172$ which are spherical in the ground state, all other
nuclei in the vicinity of the $Z=120$ line are oblate in the ground state (see Fig.\
\ref{deformation}d).  The maxima in the quantity $\delta_{2n}(Z,N)$ obtained in deformed
RHB calculations are located at completely different $Z$ values as compared with
spherical calculations indicating a possible lowering of the single-particle level density
at these values. For example, the high values of the quantity $\delta_{2p}(Z,N)$
seen at $Z=104$ around $N=150$ are due to deformed the $Z=104$ shell gap which exists
for a number of CEDFs \cite{A250,DABRS.15}.

These results clearly illustrate the danger of misinterpretation of the
structure of superheavy nuclei when using results of spherical calculations.
The presence of large spherical shell gaps will definitely manifest itself in
the increase of the relevant $\delta_{2n}(Z,N)$ or $\delta_{2n}(Z,N)$ quantities.
However, the restriction to spherical shapes does not allow to access the softness
of the potential energy surfaces and the presence of large shell gaps at deformation.

%%%%%%%%%%%%%%%%%%%%%%%%%%%%%%%%%%%%%%%%%%%%%%%%%%%%%%%%%%%%%%
\begin{figure*}[ht]
\centering
\includegraphics[angle=0,width=14cm]{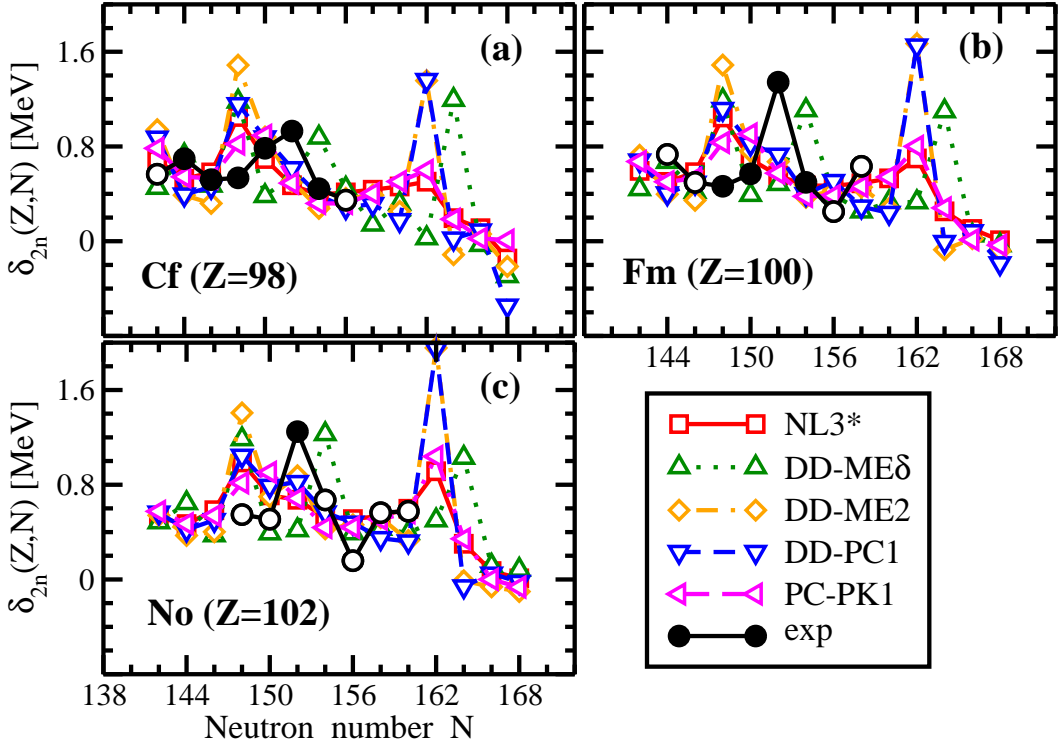}
\caption{(Color online) The quantity $\delta_{2n}(Z,N)$ for the Cf, Fm and 
No isotope chains. The experimental data (circles) are compared with the
results (open symbols) obtained in deformed RHB calculations with
the indicated CEDFs.  Solid circles are used for the
$\delta_{2n}(Z,N)$ values which are determined from measured masses and
open circles for those the definition of which involves at least
one estimated mass. The measured and estimated masses are from
Ref.\ \cite{AME2012}.}
\label{Delta_N=152}
\end{figure*}
%%%%%%%%%%%%%%%%%%%%%%%%%%%%%%%%%%%%%%%%%%%%%%%%%%%%%%%%%%%%%%%

%%%%%%%%%%%%%%%%%%%%%%%%%%%%%%%%%%%%%%%%%%%%%%%%%%%%%%%%%%%%%%
\begin{figure*}[ht]
\centering
\includegraphics[angle=0,width=14cm]{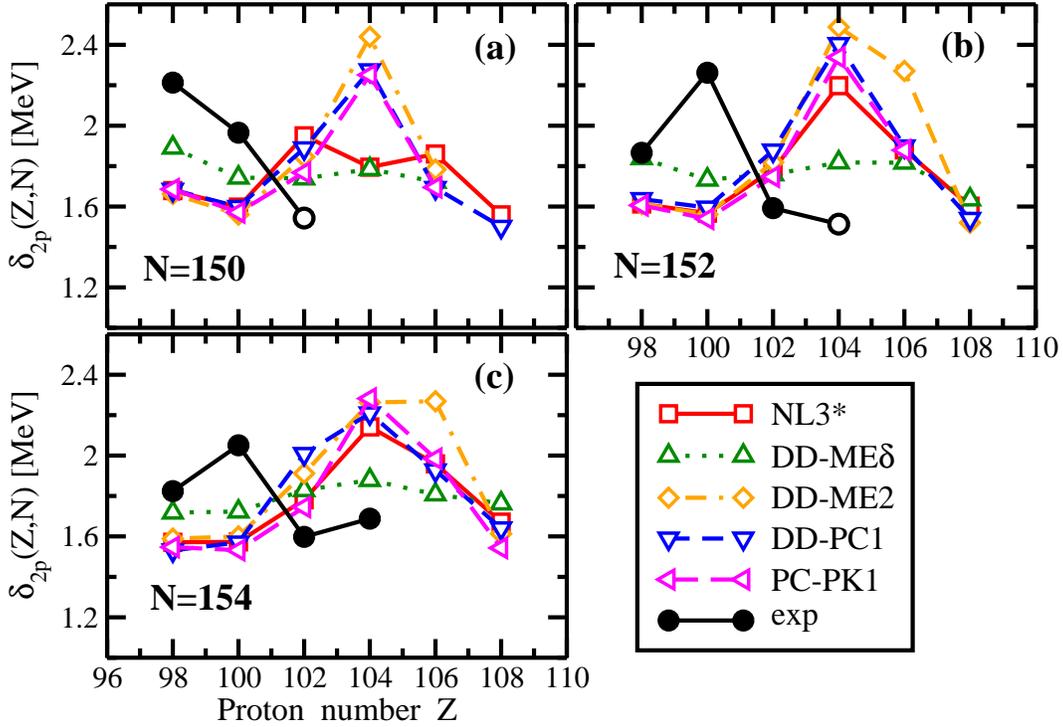}
\caption{(Color online) The same as Fig.\ \ref{Delta_N=152} but for 
the quantity $\delta_{2p}(Z,N)$ in the $N=150, 152$ and 154 isotone 
chains.}
\label{Delta_Z=100}
\end{figure*}
%%%%%%%%%%%%%%%%%%%%%%%%%%%%%5%%%%%%%%%%%%%%%%%%%%%%%%%%%%%%%%%%

%%%%%%%%%%%%%%%%%%%%%%%%%%%%%%%%%%%%%%%%%%%%%%%%%%%%%%%%%
\section{Masses and separation energies}
\label{Sep-energies}
%%%%%%%%%%%%%%%%%%%%%%%%%%%%%%%%%%%%%%%%%%%%%%%%%%%%%%%%%

In Table \ref{deviat} we list the rms-deviations $\Delta E_{\rm rms}$
between theoretical and experimental binding energies for the nuclei
with $Z\geq 96$; experimental masses from the AME2012 mass
evaluation \cite{AME2012} are used here. The masses given in
the AME2012 mass evaluation \cite{AME2012} can be separated into two
groups. One represents nuclei with masses defined only from experimental
data, the other contains nuclei with masses depending in addition on
either interpolation or short extrapolation procedures. These procedures
involve some degree of subjectivity but has proven to provide a quite
accurate estimate in absolute majority of the cases as seen from the comparison
of these estimates with newly measured masses \cite{AME2012-first}. For
simplicity, we call the masses of the nuclei in the first and second
groups as measured and estimated. Estimated masses frequently
involve the ones for unknown nuclei and they are estimated from the
trends in mass surfaces \cite{AME2012-first}. Note that these mass
surfaces also incorporate the information on odd and odd-mass SHEs
which are more abundant than their even-even counterparts \cite{OU.15}.
Experimental physical observables which depend only on measured masses
will be shown later by solid symbols in the figures, while the ones which
involve at least one estimated mass by open symbols.

  For each employed functional the accuracy of the description
of the sets of measured and measured+estimated masses is comparable
and does not change substantially when the estimated masses are added to
the measured ones (see Table \ref{deviat}). The same is true for the
quantities which depend on the mass differences such as the two-neutron
(two-proton) separation energies and the $Q_{\alpha}$ values. This
fact is important because the measured masses represent only 41\% in
the set of measured+estimated masses used here. It adds additional
support to the estimation procedures used in Refs.\ \cite{AME2012-first}
since global studies of Ref.\ \cite{AARR.14} indicate that CDFT has a
good predictive power in the regions of deformed nuclei with no shape
coexistence and the absolute majority of the superheavy nuclei for
which measured and  estimated masses are provided in Ref.\
\cite{AME2012} belong to this type of region.

  As compared with the global analysis of Refs.\ \cite{AARR.14,
ZNLYM.14}, the accuracy of the description of masses is
better for DD-PC1 and DD-ME2, comparable for DD-ME$\delta$
and PC-PK1 and worse for NL3*. The best accuracy is
achieved for DD-PC1. This is not surprising considering that this
functional has been carefully fitted to the binding energies of deformed
rare-earth nuclei and actinides in Ref. \cite{DD-PC1}. With respect to masses it
outperforms other functionals in these regions (see Figs. 6 and 7 in Ref.\
\cite{AARR.14}).

%%%%%%%%%%%%%%%%%%%%%%%%%%%%%%%%%%%%%%%%%%%%%%%%%%%%%%%%%%%%%%
\begin{figure*}[ht]
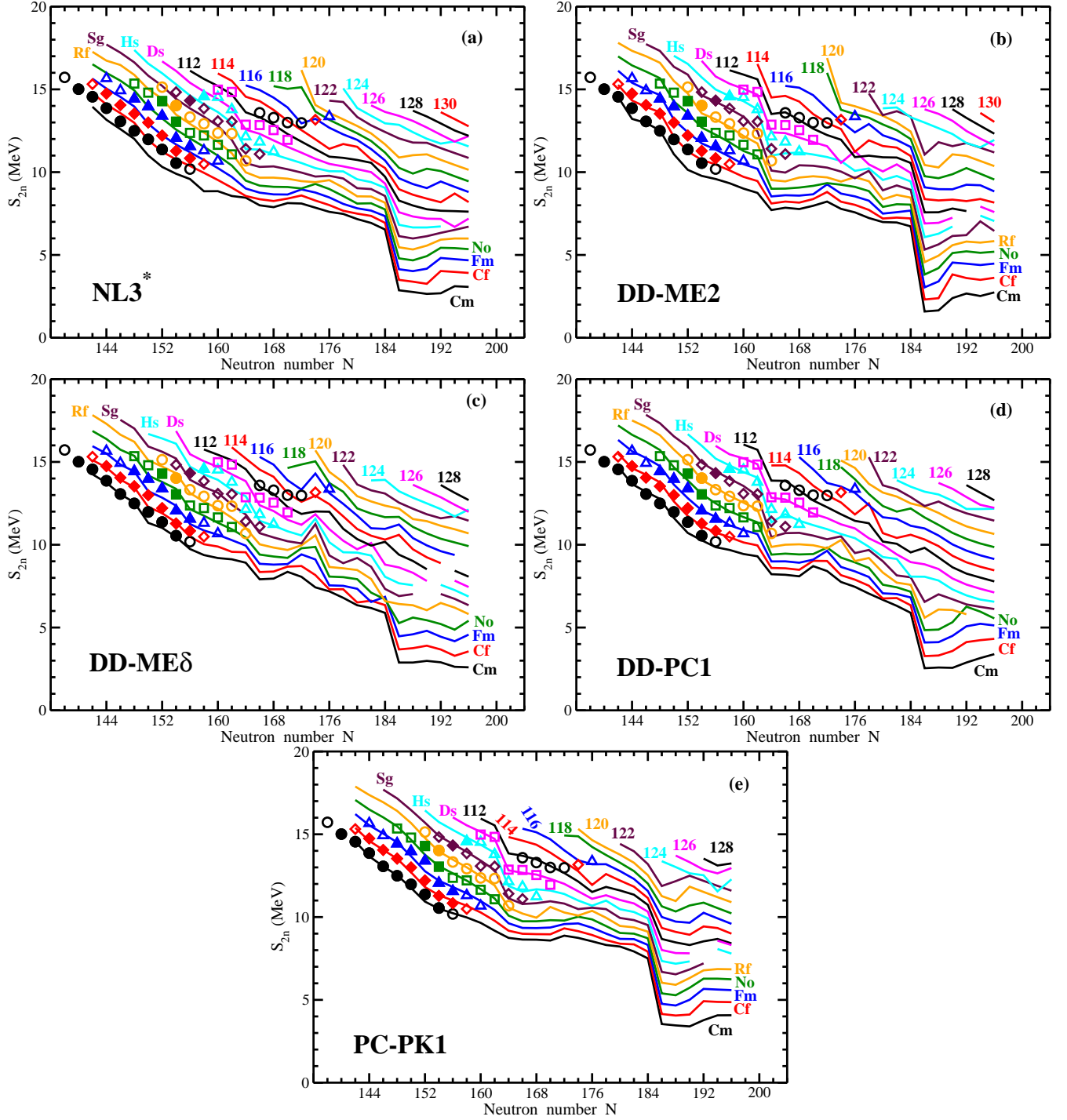

\centering
\includegraphics[angle=0,width=8.8cm]{fig-12-a.eps}
\includegraphics[angle=0,width=8.8cm]{fig-12-b.eps}
\includegraphics[angle=0,width=8.8cm]{fig-12-c.eps}
\includegraphics[angle=0,width=8.8cm]{fig-12-d.eps}
\includegraphics[angle=0,width=8.8cm]{fig-12-e.eps}
\caption{(Color online) Two-neutron separation energies $S_{2n}(Z,N)$
given for different isotopic chains as a function of neutron number.
The experimental data are shown by the symbols and the calculated results
by the lines; the same color is used for both quantities belonging
to the same isotopic chain. Solid symbols are used for the
$S_{2n}(Z,N)$ values determined from measured masses, and
open symbols for those including at least
one estimated mass. The measured and estimated masses are from
Ref.\ \cite{AME2012}. The transition to a prolate minimum with
$\beta_2 \sim 0.35$ in nuclei with $N\sim 192$ (see Fig.\
\ref{deformation}) creates the jumps in $S_{2n}(Z,N)$; these jumps are
not shown and are detectable in the graphs by the breakage of the
line.}
\label{Sep-2-neut}
\end{figure*}
%%%%%%%%%%%%%%%%%%%%%%%%%%%%%%%%%%%%%%%%%%%%%%%%%%%%%%%%%%%%%

%%%%%%%%%%%%%%%%%%%%%%%%%%%%%%%%%%%%%%%%%%%%%%%%%%%%%%%%%%%%%%
\begin{figure*}[ht]
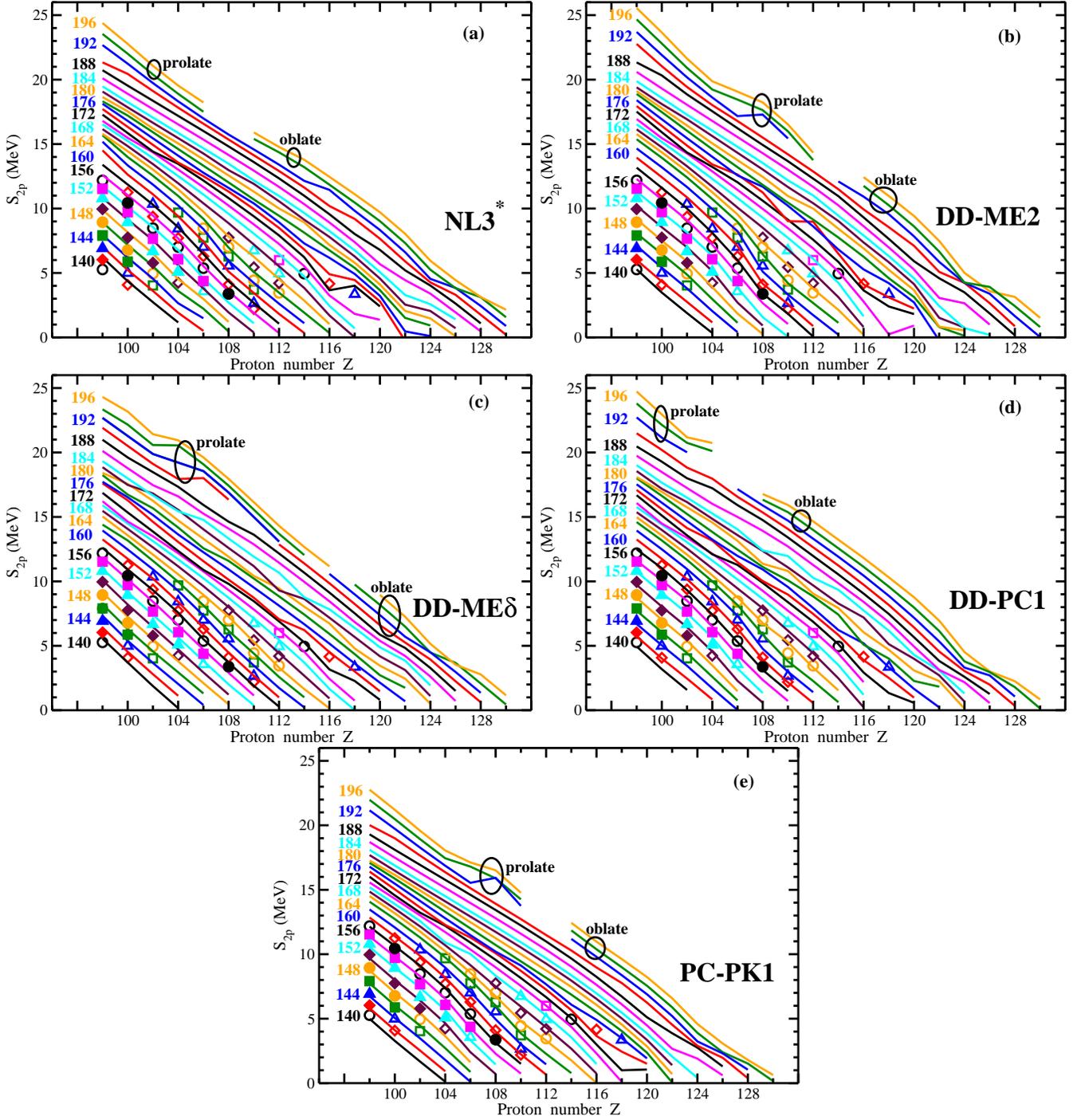

\centering
\includegraphics[angle=0,width=8.8cm]{fig-13-a.eps}
\includegraphics[angle=0,width=8.8cm]{fig-13-b.eps}
\includegraphics[angle=0,width=8.8cm]{fig-13-c.eps}
\includegraphics[angle=0,width=8.8cm]{fig-13-d.eps}
\includegraphics[angle=0,width=8.8cm]{fig-13-e.eps}
\caption{ (Color online) The same as Fig.\ \ref{Sep-2-neut} but
for two-proton separation energies $S_{2p}(Z,N)$ given for different
isotonic chains as a function of the proton number. The shapes of the
nuclei on the different sides of the $N=188-196$ chains (see Fig.\
\ref{deformation}) are indicated; the circles envelop the isotone
chains which are affected by sharp prolate-oblate transitions.
This transition creates the jumps in $S_{2p}(Z,N)$; these jumps are
not shown and are detectable in the graphs by the breakage of the
line.}
\label{Sep-2-prot}
\end{figure*}
%%%%%%%%%%%%%%%%%%%%%%%%%%%%%%%%%%%%%%%%%%%%%%%%%%%%%%%%%%%%%

  Since our investigation is restricted to even-even nuclei, we
consider two-neutron $S_{2n}=B(Z,N-2)-B(Z,N)$ and two-proton
$S_{2p}=B(Z-2,N)-B(Z,N)$ separation energies. Here $B(Z,N)$ stands
for the binding energy of a nucleus with $Z$ protons and $N$
neutrons. Apart of the proton subsystem in NL3* and DD-ME2 and
the neutron subsystem in NL3*, the two-neutron $S_{2n}$ and the two-proton $S_{2p}$
separation energies are described with a typical accuracy of 0.5 MeV
(Table \ref{deviat}). This is better by a factor of two than the
global accuracy of around 1 MeV obtained for these functionals in Ref.\
\cite{AARR.14}. The accuracy of the description of separation energies
depends on the accuracy of the description of mass differences. As a result, not
always the functional which provides the best description of masses
gives the best description of two-particle separation energies.

Figs. \ref{Sep-2-neut} and \ref{Sep-2-prot} present a detailed comparison
of calculated and experimental two-neutron and two-proton separation energies.
While providing in general comparable descriptions of experimental data,
the calculated results differ in details. The experimental data for the
Rf, Sg, Hs and Ds isotopes clearly show a sharp decrease of the two-neutron
separation energies at $N=162$ which is due to the deformed shell gap at
this particle number.  This decrease is best described
by PC-PK1 (Fig.\ \ref{Sep-2-neut}e). DD-ME2 and DD-PC1
overestimate this decrease somewhat (Figs.\ \ref{Sep-2-neut}b and d)
and NL3* underestimates its size. In contradiction to
experiment, DD-ME$\delta$ does not show the presence of a gap at
$N=162$ but gives a small  deformed shell gap at $N=164$ (Fig.\
\ref{Sep-2-neut}c).

  It is important to recognize that the conclusions about the deformed
$N=162$ shell gap are based on the comparison with two-neutron separation
energies extracted from estimated masses.
However, this gap is present both in the macroscopic+microscopic
calculations of Refs.\
\cite{PS.91,SMP.01,SP.07}  and the DFT calculations based on the Gogny D1S force
of Ref.\ \cite{WE.12}. In addition, there are indications on
the presence of this gap from the analysis of experimental data
which suggests that the deformed $N=162$ shell gap is much larger than
the $N=152$ gap discussed below \cite{AME2012-first}.

 For higher $N$ values there are indications of
the presence of the $N=184$ spherical shell gap. However, there is a substantial
difference between the functionals on how far this gap propagates into the
region of superheavy nuclei.
For example, PC-PK1 and DD-ME2 show the propagation of this gap up to
$Z\approx 120$ (Figs.\ \ref{Sep-2-neut}b and e). On the other hand, this gap is
visible only up to the Rf/No region for DD-ME$\delta$ and DD-PC1
(Figs.\ \ref{Sep-2-neut}b and e). The results for NL3* are in between of these
two extremes (Figs.\ \ref{Sep-2-neut}a).

Contrary to the neutron subsystem, experimental two-proton separation energies
are smoother as a function of proton number without clear indications
of pronounced shell gaps (Fig.\ \ref{Sep-2-prot}). One should note that
there exist small deformed shell gaps at $Z=100$ and $N=152$ in heavy
actinides/light superheavy nuclei \cite{A250}. They are barely visible
in the two-particle separation energies (see Figs.\ \ref{Sep-2-neut} and
\ref{Sep-2-prot} and Ref.\ \cite{A250}) and are usually seen in the
quantities $\delta_{2n}(Z,N)$ and $\delta_{2p}(Z,N)$ (see Sect.\
\ref{Sect-delta_2}).

Figs.\ \ref{Delta_N=152} and \ref{Delta_Z=100} show that these
deformed gaps at $Z=100$ and $N=152$ are not reproduced in the CEDFs
under consideration. Indeed, the calculations with NL3*, DD-ME2, DD-ME$\delta$,
and DD-PC1 place them at $N=148$ and $Z=104$. These gaps are clearly seen in the nobelium
region in the Nilsson diagrams for NL3* (see Fig. 3 in Ref.\ \cite{DABRS.15}).
On the contrary, a neutron gap is seen at $N=154$ and no proton gap exist
in the calculations with DD-ME$\delta$. These problems exist also in the description
of experimental deformed gaps with the older generation of the
CEDFs used in Ref. \cite{A250}. They place a neutron gap either at $N=148$
(NL3, NLRA1 and NL-Z) or at $N=150$ (NL1 and NL-Z)
or do not show a gap at all (NLSH). In the same way, a proton
gap is placed at $Z=104$ (NL3, NL1, and NL-Z) or does not
exist in NLRA1. Only NLSH predicts a proton gap at the right $Z$
value but it is placed between wrong states \cite{A250}. Note that
this problem is not specific only for covariant functionals; most
of the Skyrme functionals also fail to reproduce these gaps
(Ref.\ \cite{BRBRMG.98,SDG.14}).

%%%%%%%%%%%%%%%%%%%%%%%%%%%%%%%%%%%%%%%%%%%%%%%%%%%%%%%%%%%%%%%%%%%%%%%%%%%%%%%%%%%%
\section{$\alpha$-decay properties}
\label{Alpha-decay}
%%%%%%%%%%%%%%%%%%%%%%%%%%%%%%%%%%%%%%%%%%%%%%%%%%%%%%%%%%%%%%%%%%%%%%%%%%%%%%%%%%%%

In superheavy nuclei spontaneous fission and $\alpha$ emission compete
and the shortest half-live determines the dominant decay channel  and the total
half-live. Only in cases where the spontaneous fission half-live
is longer than the half-live of $\alpha$ emission superheavy nuclei can be
observed in experiment. In addition, only nuclei with half-lives longer than
$\tau =10\mu$s are observed in experiments.

The $\alpha$ decay half-live depends on the $Q_{\alpha}$ values which are
calculated  according to
\begin{equation}
Q_{\alpha}=E(Z,N)-E(Z-2,N-2)-E(2,2)
\end{equation}
with $E(2,2)=-28.295674$ MeV \cite{AME2012} and $Z$ and $N$ representing
the parent nucleus.

\begin{figure*}[ht]
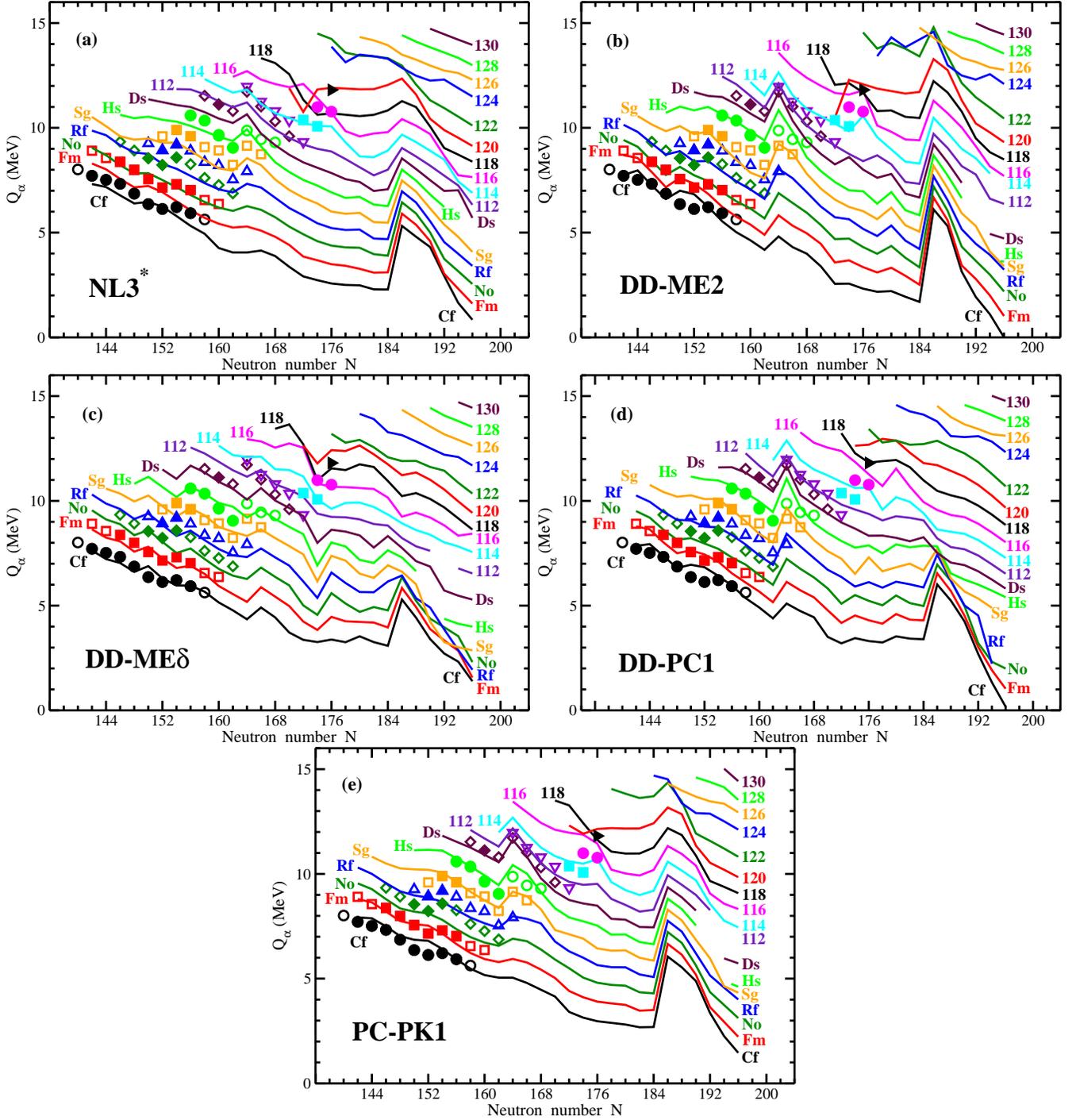

\centering
\includegraphics[angle=0,width=8.8cm]{fig-14-a.eps}
\includegraphics[angle=0,width=8.8cm]{fig-14-b.eps}
\includegraphics[angle=0,width=8.8cm]{fig-14-c.eps}
\includegraphics[angle=0,width=8.8cm]{fig-14-d.eps}
\includegraphics[angle=0,width=8.8cm]{fig-14-e.eps}
\caption{ (Color online) The comparison of experimental and
calculated $Q_{\alpha}$ values for even-even superheavy nuclei.
The experimental and calculated values are shown by symbols
and lines, respectively. For a given isotope chain, the same
color is used for both types of values. Experimental $Q_{\alpha}$
values are from Ref.\ \protect\cite{AME2012}. Solid symbols
are used for experimentally measured $Q_{\alpha}$ values \cite{AME2012}
which are  determined either from measured masses (for low-$Z$ values)
or from $\alpha$-decays (for high-$Z$ values). Open symbols are
used for the $Q_{\alpha}$ values the determination of which involves
at least one estimated mass.}
\label{Q_alpha}
\end{figure*}

  The RHB results for the $Q_{\alpha}$ values are compared with experiment in Fig.\
\ref{Q_alpha} and the corresponding rms-deviations are listed in Table \ref{deviat}.
Based on the results presented in this table, the best agreement is obtained
for PC-PK1 closely followed by DD-PC1 and DD-ME$\delta$, and then by DD-ME2 and NL3*.
However, a detailed analysis of these results presented in Fig.\
\ref{Q_alpha} clearly indicates that DD-ME$\delta$ completely
misses both the position in neutron number and the magnitude of the
peak at $N=164$ seen in the experimental data for the Rf, Sg, Hs, and
Ds isotope chains.  Note, however, that the magnitude of the
peak in the experimental data is based on the estimated masses. This peak is a
consequence of the deformed $N=162$ shell gap which is not reproduced in
this functional (see Sec.\ \ref{Sep-energies}). The other functionals correctly
place this peak at $N=164$. The best reproduction of the magnitude of this
peak is obtained for PC-PK1.  The CEDFs DD-PC1 and DD-ME2 (NL3*) somewhat
overestimate (underestimate) its magnitude reflecting the accuracy of the
reproduction of the size of the  $N=162$ shell gap in these
CEDFs (see Sec.\ \ref{Sep-energies}).

%%%%%%%%%%%%%%%%%%%%%%%%%%%%%%%%%%%%%%%%%%%%%%%%%%%%%%%%%%%%%%
%%%%%%%%%%%%%%%%%%%%%%%%%%%%%%%%%%%%%%%%%%%%%%%%%%%%%%%%%%%%%

%%%%%%%%%%%%%%%%%%%%%%%%%%%%%%%%%%%%%%%%%%%%%%%%%%%%%%%%%%%%%%
\begin{figure*}[ht]
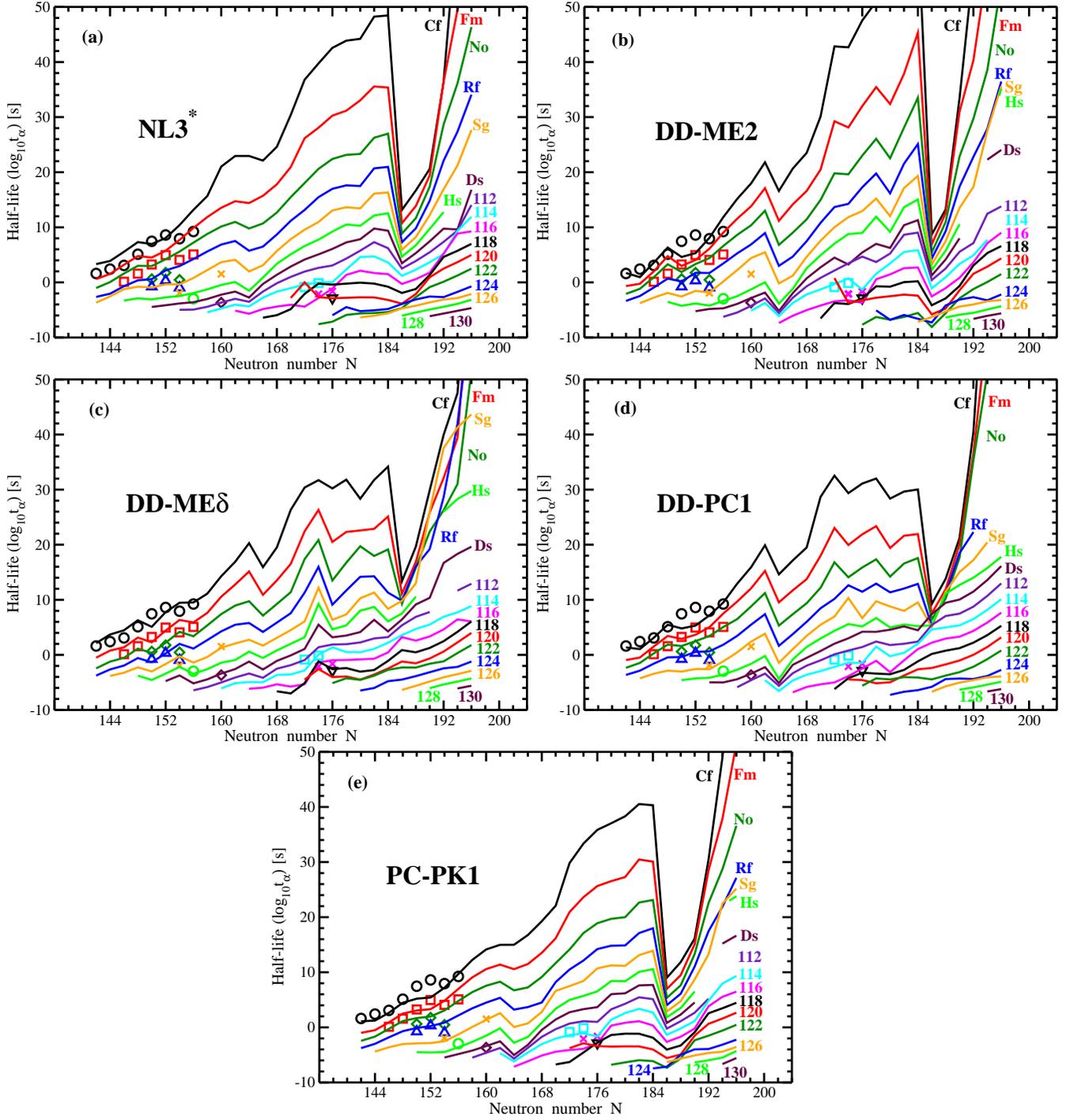

\centering
\includegraphics[angle=0,width=8.8cm]{fig-15-a.eps}
\includegraphics[angle=0,width=8.8cm]{fig-15-b.eps}
\includegraphics[angle=0,width=8.8cm]{fig-15-c.eps}
\includegraphics[angle=0,width=8.8cm]{fig-15-d.eps}
\includegraphics[angle=0,width=8.8cm]{fig-15-e.eps}
\caption{(Color online) Experimental and calculated half-lives
for $\alpha$-decays of even-even superheavy nuclei. The experimental
and calculated values are shown by symbols and lines, respectively.
For a given isotope chain, the same color is used for both types
of values. The experimental data are from Ref.\
\cite{AME2012-decay}.}
\label{Q_alpha-decay}
\end{figure*}
%%%%%%%%%%%%%%%%%%%%%%%%%%%%%%%%%%%%%%%%%%%%%%%%%%%%%%%%%%%%%

The comparison of experimental data with theoretical $Q_{\alpha}$
values obtained with the covariant functionals (Fig.\ \ref{Q_alpha} of the present
manuscript and Fig. 18 of Ref.\ \cite{BHP.03}), and with those obtained
by non-relativistic models (see, for example, Fig. 18 in Ref.\
\cite{BHP.03} and Figs. 44 and 45 of Ref.\ \cite{SP.07}) clearly
indicates that the available experimental data do not allow to
distinguish the predictions of different models with respect to
the position of the center of the island of stability.

The $\alpha$-decay half-lives were computed using the phenomenological
Viola-Seaborg formula \cite{VS.66}
\begin{equation}
log_{10}\tau_{\alpha}=\frac{aZ+b}{\sqrt{Q_{\alpha}}}+cZ+d
\end{equation}
with the parameters $a=1.64062$, $b=-8.54399$,
$c=-0.19430$ and $d=-33.9054$ of Ref. \cite{DR.05}.

The comparison of calculated and experimental half-lives for
the $\alpha$-decays is presented in Fig.\ \ref{Q_alpha-decay}. One
can see that reasonable agreement is obtained for all functionals
especially for the case of PC-PK1. However, the
local increase above the general trend of the experimental
half-lives near $N=152$ visible in the Cf, Fm and No isotope
chains, which is due to deformed $N=152$ shell gap, is not
reproduced. Neither of the functionals reproduce the
position of this gap (see Sec.\ \ref{Sep-energies}). For
higher neutron numbers all functionals
predict an increase of the half-lives as a function of
neutron number $N$. This trend is however interrupted in the
vicinity of the spherical shell gap with $N=184$. For some isotope
chains a drastic decrease of the half-lives is
observed. It is a consequence of the well known fact that
for nuclei with two neutrons outside a closed shell
$\alpha$-particle emission is easier than for
the other nuclei in the same isotopic chain \cite{BZZRL.14}. However,
above $N=184$ the trend of increasing half-lives with
the increase of neutron number is restored. The impact of the
$N=184$ shell gap on the $\alpha$-decay half-lives clearly
correlates with the impact of this gap of the deformations
of the ground states (Sec.\ \ref{def-system}). In SHEs with
high $Z$ values its impact on the $\alpha$-decay half lives
is either substantially decreased or completely vanishes.

In the region under investigation the magnitude of the $\alpha$ decay
half-lives varies in a very wide range from $10^{-8}$ up
to $10^{50}$ s (or even higher for  the Cf, Fm and No nuclei
with $N\sim 190$). For some SHEs with high-$Z$ values the calculated
half-lives fell below the experimental observation limit
of $10^{-5}$s.

Despite the fact that the existing experimental data on the
$\alpha$-decay half-lives is described with comparable
accuracy by the different functionals, for unknown
regions of nuclear chart there are some cases of
substantial difference in their predictions. The most extreme difference is
seen in the Cf isotopes, where NL3* and DD-ME2 differ from DD-ME$\delta$ and
DD-PC1 by approximately 20 orders of magnitude at neutron
number $N=184$ and slightly below it. On the other hand,
apart from the $N=184$ region the differences in the
predictions of different functionals
is smaller for SHEs with $Z\sim 114$ where it reaches only
few orders of magnitude (Fig.\ \ref{Q_alpha-decay}). In the
$N=184$ region of these nuclei the differences between
predictions of different functionals increase by additional
few orders of magnitude. However, above $Z=120$ these differences
decrease with increasing proton number because of the diminishing
role of the $N=184$ spherical shell gap. For example, the
differences in the predicted $\alpha$-decay half-lives do not
exceed two orders of magnitude for the $Z=128$ and 130 nuclei.

%%%%%%%%%%%%%%%%%%%%%%%%%%%%%%%%%%%%%%%%%%%%%%%%%%%%%%%
\section{Conclusions}
\label{concl}
%%%%%%%%%%%%%%%%%%%%%%%%%%%%%%%%%%%%%%%%%%%%%%%%%%%%%%%

The performance of covariant energy density functionals
in the region of superheavy nuclei has been assessed
using the state-of-the-art functionals NL3*, DD-ME2, DD-ME$\delta$,
DD-PC1, and PC-PK1. They represent major classes of
covariant functionals with different basic model assumptions
and fitting protocols. The available experimental data on
ground state properties of even-even nuclei have been confronted
with the results of the calculations. For the first time,
theoretical spreads in the prediction of physical observables have been
investigated in a systematic way in this region of the nuclear chart
for covariant density functionals. Special attention has been paid to
the propagation of these spreads towards unknown regions of
higher $Z$ values and of more neutron-rich nuclei.

 The main results of this work can be summarized as follows:
\begin{itemize}

\item
So far, the absolute majority of investigations of the shell
structure of SHEs has been performed in spherical calculations. In the
framework of covariant density functional theory, these calculations always
indicate a large proton shell gap at $Z=120$, a smaller neutron shell gap at $N=172$,
and, for some functionals, a neutron shell gap at $N=184$. However, the restriction to
spherical shapes does not allow to access the softness of the potential
energy surfaces and the presence of competing large shell gaps at
deformation. As illustrated in the present manuscript, this restriction
has led to a misinterpretation of the shell structure of SHEs. The detailed
analysis, with deformation included, shows that the impact of the $N=172$
shell gap is very limited in the $(Z,N)$ space for all functionals under investigation.
The impact of the $Z=120$ and $N=184$ spherical shell gaps depend drastically
on the functional. It is most pronounced for NL3* and PC-PK1 and is (almost) completely absent
for DD-PC1 and DD-ME$\delta$.

\item
Available experimental data (separation energies, $Q_{\alpha}$-values
and $\alpha$-decay half-lives) on SHEs are described with
comparable accuracy in covariant (current manuscript) and non-relativistic
\cite{BHP.03} DFT calculations. Moreover, these data are not
very sensitive to the details of the single-particle structure which
define the position of the center of the island of stability.
Unfortunately, experimental  data on single-particle states in
odd-mass SHEs are either not available ($Z\geq 105$) or
scarce ($100<Z<105$). In addition, in the latter case, the configuration
assignments for many nuclei are not fully reliable \cite{255Lr}.  However,
the analysis of the available data on deformed single-particle states in the
actinides performed with different DFTs in Refs.\ \cite{AS.11,DABRS.15} reveals
that the problems in the description of the single-particle structure exist
in all models. Considering the existing theoretical spreads, it is
clear that the available experimental data on SHEs do not allow to distinguish
the predictions of different models with respect of the position of the center
of the island of stability.

\item
Comparing different functionals one can see that the results
obtained with the covariant density functional DD-ME$\delta$  differ
substantially from the results of other functionals. This functional is
different from all the other functionals used here, because it has been
adjusted in Ref. \cite{DD-MEdelta} using only four phenomenological
parameters in addition to some input from ab-inito calculations
\cite{Baldo2004_NPA736-241,VanDalen2007_EPJA31-29}. Above $Z=102$ it
does not predict spherical
SHEs. The heights of the inner fission barriers in SHEs with $Z=112-116$
obtained in this functional are significantly lower than the experimental
estimates and the values calculated in all other models. In addition, it
does not lead to octupole deformation in actinide nuclei which
are known to be octupole deformed \cite{AAR.14}.
All these facts suggest that either the ab-initio input \cite{Baldo2004_NPA736-241,VanDalen2007_EPJA31-29}
for this functional is not precise enough or the number of only four phenomenological
parameters (fitted to masses of spherical nuclei) is too small to provide a
proper extrapolation to the region of superheavy elements.
Thus, this functional is not recommended for future investigations in this area,
in spite of the fact that this functional provides a good description of masses
and other ground state observables in the $Z\leq 104$ nuclei \cite{AARR.14}.

\item
Theoretical uncertainties in the predictions of different observables
have been quantified. While the uncertainties in the quadrupole
deformation of the ground states of known superheavy nuclei are small,
they increase on approaching nuclei with $Z=120$ and/or $N=184$. As a result,
even the ground state deformations of these nuclei (whether spherical
or oblate) cannot be predicted with certainty. Available experimental
data do not allow to discriminate between these predictions. 
%Quite
%substantial theoretical {\bf spreads} exist also for the inner fission
%barrier heights. It is clear that they are not constrained well
%by the current fitting protocols based on nuclear matter and on
%bulk properties of a very restricted set of nuclei.
%However, it was shown that benchmarking of the functionals to the experimental
%data on fission barriers in the actinides allows to reduce the theoretical
%uncertainties for the inner fission barriers of unknown nuclei.
%Even then the trend of increasing
%theoretical uncertainties in the neutron-rich SHEs persists. This will
%most likely have a profound impact not only on our understanding of SHEs
%but also on the r-process simulations of neutron-star mergers.

\end{itemize}

\begin{acknowledgments}
 This material is based upon work supported by the U.S. Department of
Energy, Office of Science, Office of Nuclear Physics under Award
Number DE-SC0013037 and by the DFG cluster of excellence
\textquotedblleft Origin and Structure of the Universe\textquotedblright\
(www.universe-cluster.de). A.A. and T.N. thank the JSPS invitation
fellowship in Japan (S-15029) for financial support during completion
of the present work.  It is also partially supported by JSPS KAKENHI
(Grants No. 24105006 and 25287065)) and by ImPACT Program of
Council for Science, Technology and Innovation (Cabinet Office,
Government of Japan).
\end{acknowledgments}

%%%%%%%%%%%%%%%%%%%%%%%%%%%%%%%%%%%%%%%%%%%%%%%%%%%%%%%%%%%%%%%%
\appendix*
\section{Supplemental information on the ground state
properties}
%%%%%%%%%%%%%%%%%%%%%%%%%%%%%%%%%%%%%%%%%%%%%%%%%%%%%%%%%%%%%%%%

     In addition to the graphical representation of the results, the numerical
results for ground state properties obtained with the DD-PC1 and PC-PK1 CEDFs
are provided in two tables of the Supplemental Material with this article
as Ref.\ \cite{Sup}. The structure of these tables is illustrated in Table
\ref{sup_table}.

%%%%%%%%%%%%%%%%%%%%%%%%%%%%%%%%%%%%%%%%%%%%%%%%%%%%%%%%%%%%%%%%%%%%%%%%%%%%%%%%%%%%%%%%%%
\begin{table*}[htbp]
\caption{The RHB predictions for ground state properties of even-even nuclei
obtained with PC-PK1. Columns 3, 4 and 5 list binding energies $E$,
proton ($\beta_{p}$) and neutron ($\beta_{p}$) quadrupole deformations.
The charge radii $r_{ch}$, root-mean square (rms) proton radii $r_{rms}^p$
and neutron skin thicknesses  $r_{skin}$ are presented in columns 6, 7
and 8. The following notation $r_{rms}^p=<r_p^2>^{1/2}$ is used. Note that the
neutron root-mean-square (rms) radii can be calculated as $r_{rms}^n=r_{rms}^p+r_{skin}$.
The last column gives alpha-decay half-lives obtained by means of the
Viola-Seaborg formula (see Sec.\ \ref{Alpha-decay} for details.)
}
\begin{center}
\begin{tabular}{|c|c|c|c|c|c|c|c|c|} \hline
 $Z$  &  $N$  & $E$(MeV) &  $\beta_{p}$  &  $\beta_{n}$  & $r_{ch}$ (fm) &  $r_{rms}^p$(fm)  & $r_{skin}$(fm)  &  $log_{10}(t_{\alpha})$(s) \\
 \hline
 1   &   2   &      3      &     4    &     5    &    6    &     7    &     8    &     9  \\ \hline
118  &  170  &  -2034.295  &   0.168  &   0.163  &  6.287  &   6.236  &   0.150  &   -6.731 \\
118  &  172  &  -2049.224  &   0.000  &   0.000  &  6.271  &   6.220  &   0.153  &   -6.287 \\
118  &  174  &  -2064.100  &   0.000  &   0.000  &  6.278  &   6.227  &   0.166  &   -4.563 \\
118  &  176  &  -2078.333  &   0.000  &   0.000  &  6.285  &   6.233  &   0.180  &   -2.755 \\
118  &  178  &  -2092.081  &   0.000  &   0.000  &  6.291  &   6.240  &   0.192  &   -1.385 \\
118  &  180  &  -2105.365  &   0.000  &   0.000  &  6.297  &   6.246  &   0.206  &   -1.147 \\
118  &  182  &  -2118.141  &   0.000  &   0.000  &  6.302  &   6.251  &   0.221  &   -1.152 \\
118  &  184  &  -2130.160  &   0.000  &   0.000  &  6.307  &   6.256  &   0.236  &   -1.873 \\
118  &  186  &  -2140.809  &   0.000  &   0.000  &  6.329  &   6.278  &   0.241  &   -4.035 \\
118  &  188  &  -2151.145  &   0.000  &   0.000  &  6.350  &   6.299  &   0.246  &   -3.289 \\
118  &  190  &  -2161.846  &  -0.400  &  -0.395  &  6.603  &   6.555  &   0.223  &   -0.900 \\
118  &  192  &  -2172.715  &  -0.407  &  -0.403  &  6.623  &   6.574  &   0.234  &    2.559 \\
118  &  194  &  -2183.257  &  -0.413  &  -0.411  &  6.643  &   6.595  &   0.244  &    3.462 \\
118  &  196  &  -2193.479  &  -0.420  &  -0.419  &  6.666  &   6.617  &   0.253  &    4.458 \\ \hline
\end{tabular}
\label{sup_table}
\end{center}
\end{table*}

\bibliography{references9}

\end{document}